\documentclass[
 aps,prb,
 amsmath,amssymb,
 reprint,%
 showpacs,superscriptaddress,
]{revtex4-2}
\usepackage{bm,amsfonts,amssymb,amsmath,mathrsfs}
\usepackage[colorlinks,linkcolor=blue,citecolor=blue,urlcolor=blue]{hyperref}

\usepackage{lineno}
\usepackage{lipsum}

\usepackage{xcolor}
\usepackage{epsfig}
\usepackage{graphicx}
\usepackage{dcolumn}
\usepackage{braket}
\usepackage{bbold}

\newcommand{\beq}{\begin{equation}}
\newcommand{\eeq}{\end{equation}}
\newcommand{\beqa}{\begin{eqnarray}}
\newcommand{\eeqa}{\end{eqnarray}}

\makeatletter
\newcommand*{\revision}{\@ifnextchar\bgroup{ @}{\color{black}}}
\newcommand*{\revision@}[1]{{\textcolor{black}{#1}}}
\makeatother

\begin{document}

\title{Coherent-Response and Thermal-Fluctuation Information in Cavity-Magnon Magnetometry}

\author{Guogan Zhao}
\affiliation{Department of Physics, and Center for Quantum Science and Engineering, Stevens Institute of Technology, Hoboken, NJ 07030, USA}
\author{Xiao-Feng Qian$^*$}
\email{xqian6@stevens.edu}
\affiliation{Department of Physics, and Center for Quantum Science and Engineering, Stevens Institute of Technology, Hoboken, NJ 07030, USA}

\date{\today}

\begin{abstract}
Magnetic sensing with cavity--magnon systems is commonly optimized through coherent response functions, while field-dependent fluctuations are treated primarily as measurement noise. Here, we show that this description omits a distinct and experimentally accessible source of magnetic-field information: the field dependence of the measured quadrature variance. For repeated phase-sensitive measurements of a calibrated microwave quadrature, we derive the complete Gaussian likelihood and demonstrate that its Fisher information separates exactly into a coherent-displacement contribution,
determined by the field derivative of the quadrature mean, and a thermal-fluctuation contribution, determined by the field derivative of its variance. The two contributions arise from the same measurement record but obey different optimization principles. For the amplitude quadrature, the dimensionless coherent morphology is globally maximized at cooperativity $C=2$, corresponding to coupling strength $g=\sqrt{2\kappa_a\kappa_m}$, and at linewidth-normalized detunings $\delta_a=\delta_m=\pm1$. With a near-resonance expansion, the reduced variance-information is globally maximized at resonant cavity--magnon frequency matching, $\Delta_a=\Delta_m$, and approaches a finite large-cooperativity limit. The competing nature of the two sources of contribution is shown in detail under all parameter regions of driving frequency and temperature. Most notably, thermal variance contribution remains finite when coherent driving information is buried in noise (e.g., at low-temperature and low-excitation cases) or even vanishes. It reveals that thermal fluctuations can carry operational magnetic-field information even when the coherent-displacement contribution is unobservable or absent. These results establish a measurement-specific metrological framework and provide distinct operating rules for exploiting coherent response, thermal fluctuations, or both in cavity--magnon magnetometry.
\end{abstract}

\maketitle

\section{Introduction}

Coherent coupling between collective magnon excitations and microwave-cavity photons has established cavity magnonics as a versatile platform for hybrid light--matter physics.\cite{li2020hybrid,zarerameshti2022cavity}
Strong coupling and normal-mode splitting have been demonstrated in planar and three-dimensional resonators,\cite{huebl2013high,tabuchi2014hybridizing,zhang2014strongly,
goryachev2014highcooperativity,bai2015spin}
and cavity magnons have been interfaced with superconducting qubits, mechanical modes, optical fields and electronic circuits.\cite{tabuchi2015coherent,hisatomi2016bidirectional,osada2016cavity,
zhang2016optomagnonic,zhang2016cavity,haigh2016tripleresonant,klingler2016combined,
braggio2017optical,li2018magnonphotonphonon,osada2018brillouin,
lachance-quirion2019hybrid,li2019strong,luo2021nonlocal,shen2022coherent,
zhang2025gain,gui2026gaindriven}
For magnetic sensing, a longitudinal field shifts the magnon resonance, while magnon--photon hybridization converts this shift into a microwave response accessible through transmission, reflection, or phase-sensitive quadrature measurements.\cite{flower2019experimental,crescini2020cavity,crescini2021phasemodulated}

Previous cavity--magnon sensing studies have focused mainly on coherent observables, including transmission amplitude, phase response, dispersive shifts, and hybrid-mode spectra.
\cite{cao2019exceptional,wolski2020dissipationbased,
crescini2021phasemodulated,ebrahimi2021singlequadrature}
Such observables characterize magnetic susceptibility, but a large response does not by itself determine estimation precision, which depends on the complete probability distribution of a specified measurement record.
\cite{langbein2018no,rumyantsev2019discrete}
Although information-theoretic methods have been applied to cavity--magnon metrology,\cite{wan2024quantumenhanced} thermal fluctuations in quadrature measurements are generally treated only as noise that degrades the coherent signal.\cite{ebrahimi2021singlequadrature}
The magnetic-field information encoded in the field dependence of these fluctuations has therefore not been isolated from the complete quadrature likelihood.

Here, we develop an operational Fisher-information framework for repeated phase-sensitive measurements of a calibrated microwave quadrature. For linear cavity--magnon dynamics with Gaussian thermal inputs, both the quadrature mean and variance depend on the magnetic field. {  The resulting Fisher information separates exactly into coherent-displacement and thermal-fluctuation contributions. These are governed by the field derivatives of the mean and variance, respectively, and are contained in the same measurement record. The thermal contribution originates from the field-induced shift of the magnon frequency. This shift changes the magnon excitation energy and its thermal occupation at fixed temperature. It also changes the cavity--magnon frequency mismatch, modifying the relative weights of the two thermal reservoirs in the cavity variance. Both effects make the quadrature distribution width sensitive to the magnetic field. Even when the thermal occupations coincide at resonance, the magnon occupation retains a finite field derivative at nonzero temperature. The distribution width can therefore carry field information even without a coherent displacement. Thermal fluctuations consequently play a dual role: they suppress coherent-displacement information through broadening and provide additional information through their field-dependent variance.}

{Three principal results follow. First, coherent displacement and thermal fluctuations are identified as two operational information carriers within the same quadrature record. Second, they favor different operating conditions: for the amplitude quadrature considered here, the coherent morphology is optimized by linewidth-normalized detuning matching at cooperativity $C=2$, whereas the reduced thermal morphology is optimized by absolute cavity--magnon frequency matching within the near-resonance expansion. Third,  when the coherent signal is buried in thermal noise, the magnetic-field dependence of the thermal fluctuations provides useful information. In the weak-drive regime, including single-photon coherent driving (e.g., for quantum magnonic interaction with a superconducting qubit \cite{tabuchi2015coherent}), the variance contribution can be comparable to or even exceed the coherent-displacement contribution. At finite temperature, this information channel remains nonzero even without coherent driving. Our results reveal the fundamental role of fluctuations as carriers of magnetic-field information and provide guidance for optimizing sensing performance across different experimental regimes.}

\section{Operational Model for Cavity--Magnon Magnetometry}

\subsection{Cavity--Magnon System and Measurement Statistics}

\begin{figure}[htbp]
\centering
\includegraphics[width=\linewidth]{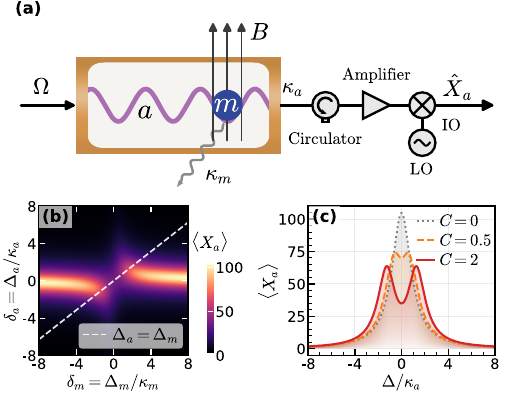}
\caption{Steady-state coherent cavity response and cavity--magnon spectral hybridization. (a) Schematic diagram. {A microwave cavity mode $\hat a$, coherently driven with amplitude $\Omega$, is coupled to a Kittel magnon mode $\hat m$ whose resonance frequency is tuned by the longitudinal magnetic field $B$. The cavity and magnon modes have damping rates $\kappa_a$ and $\kappa_m$, respectively. The cavity output is routed through a circulator and amplifier and then mixed with a local oscillator for phase-sensitive detection of the cavity amplitude quadrature $\hat X_a$. The field dependence of the resulting quadrature statistics allows the estimation of its strength via Fisher-information analysis.} (b) Mean cavity amplitude quadrature $\langle \hat X_a\rangle$ in the normalized-detuning plane $(\delta_m,\delta_a)$ at $C=2$. The avoided crossing reflects the hybridization of the cavity and magnon modes, and the white dashed line marks the symmetric absolute-detuning condition $\Delta_a=\Delta_m$. (c) Cross sections along $\Delta_a=\Delta_m\equiv\Delta$ for increasing coupling strength. The response evolves from a single resonance into two polariton branches whose characteristic detunings are determined by the real parts of the coupled-system poles. The other parameters are adopted to be: $\kappa_a/2\pi=1.35$ MHz, $\kappa_m/2\pi=1.06$ MHz and $\Omega/2\pi=100$ MHz.}
\label{fig1}
\end{figure}

We consider the cavity--magnon sensing configuration illustrated in Fig.~\ref{fig1}(a). A microwave-cavity mode with annihilation operator $\hat a$ and resonance frequency $\omega_a$ is coherently coupled to a Kittel magnon mode with annihilation operator $\hat m$ and field-dependent resonance frequency $\omega_m(B)$. \cite{zhao2021driven,zhao2025theory,liu2026quantum} A longitudinal magnetic field $B$ shifts the magnon resonance, while the cavity is coherently driven at frequency $\omega_d$ with amplitude $\Omega$, {  and the two modes interact with coupling strength $g$. The drive amplitude is related to the incident power $P_{\rm in}$ at the cavity input port by $\Omega=\sqrt{2\kappa_{\rm ex}P_{\rm in}/(\hbar\omega_d)}$, where $\kappa_{\rm ex}$ is the contribution of the driven port to the cavity amplitude decay rate and $\hbar$ is the reduced Planck constant.} The cavity output is routed through a phase-sensitive microwave readout chain, which provides access to the cavity amplitude quadrature $\hat X_a$ considered below. The corresponding Hamiltonian is
\begin{equation}
\begin{aligned}
H=&
\hbar\omega_a \hat a^\dagger \hat a
+\hbar\omega_m(B)\hat m^\dagger \hat m
+\hbar g\left(\hat a^\dagger \hat m+\hat a \hat m^\dagger\right) \\
&+i\hbar\Omega
\left(\hat a^\dagger e^{-i\omega_d t}-\hat a e^{i\omega_d t}\right).
\end{aligned}
\label{eq:hamiltonian}
\end{equation}
The longitudinal magnetic field is encoded through
\begin{equation}
\begin{aligned}
&\omega_m(B)=\gamma B, \\
&\Delta_a=\omega_d-\omega_a, \\
&\Delta_m(B)=\omega_d-\omega_m(B),
\end{aligned}\label{eq:field_encoding}
\end{equation}
where $\gamma$ is the gyromagnetic ratio, while $\Delta_a$ and $\Delta_m$ are the cavity and magnon detunings, respectively. We consider local estimation of small field variations around a chosen operating point and assume that the variation is sufficiently slow for the coupled response to follow the instantaneous steady state within the sensing bandwidth.

In the frame rotating at the drive frequency, the coupled Heisenberg--Langevin equations are
\begin{align}
\dot{\hat a}
&=\left(i\Delta_a-\kappa_a\right)\hat a
-ig\hat m+\Omega
+\sqrt{2\kappa_a}\,\hat a^{\mathrm{in}},
\label{eq:cavity_langevin}
\\
\dot{\hat m}
&=\left(i\Delta_m-\kappa_m\right)\hat m-ig\hat a +\sqrt{2\kappa_m}\,\hat m^{\mathrm{in}},
\label{eq:magnon_langevin}
\end{align}
where $\kappa_a$ and $\kappa_m$ are amplitude-decay rates. For the cavity, $\kappa_a=\kappa_{\rm ex}+\kappa_{\rm int}$ contains the coupling rate $\kappa_{\rm ex}$ to the monitored microwave port and the internal loss rate $\kappa_{\rm int}$. The operators $\hat a^{\mathrm{in}}$ and $\hat m^{\mathrm{in}}$ denote independent Gaussian Markovian inputs in the effective local-damping model.\cite{gardiner1985input,clerk2010introduction,aspelmeyer2014cavity,blais2021circuit} For either mode $\hat j\in\{\hat a,\hat m\}$, the nonzero noise correlations are
\begin{align}
\left\langle
\hat j^{\mathrm{in}\dagger}(t)\hat j^{\mathrm{in}}(t')
\right\rangle
&=n_j^{\mathrm{th}}\delta(t-t'),
\\
\left\langle
\hat j^{\mathrm{in}}(t)\hat j^{\mathrm{in}\dagger}(t')
\right\rangle
&=\left(n_j^{\mathrm{th}}+1\right)\delta(t-t'),
\label{eq:noise_correlations}
\end{align}
with thermal occupations
\begin{equation}
n_j^{\mathrm{th}} = \frac{1}{\exp[\hbar\omega_j/(k_BT)]-1}.
\label{eq:thermal_occupations}
\end{equation}
\textcolor{black}{For later comparison of the cavity and magnon responses on their respective linewidth scales, we use the normalized detunings}
\begin{equation}
\delta_a=\frac{\Delta_a}{\kappa_a},
\qquad
\delta_m=\frac{\Delta_m}{\kappa_m}.
\label{eq:normalized_detunings}
\end{equation}

The drive amplitudes considered below are assumed to remain in the regime where magnon Kerr nonlinearities are negligible. With the steady-state magnon amplitude $\beta=-ig\Omega/\lambda$, this requires the drive-induced Kerr shift to satisfy\cite{li2018magnonphotonphonon}
\begin{equation}
|K_m|\,|\beta|^3
=|K_m|\frac{g^3\Omega^3}{|\lambda|^3}
\ll \Omega,
\label{eq:linear_response_condition}
\end{equation}
where $K_m$ is the single-magnon Kerr coefficient. Stronger driving would require nonlinear corrections to Eqs.~\eqref{eq:cavity_langevin} and \eqref{eq:magnon_langevin}. \cite{wang2016magnon,zhang2019theory,zheng2023tutorial}

Writing the steady-state cavity amplitude as $\alpha=\langle\hat a\rangle$, the coherent solution is
\begin{equation}
\alpha
=\frac{\Omega\lambda_m}{\lambda}.
\label{eq:cavity_amplitude}
\end{equation}
where the complex rates are $\lambda_a=\kappa_a-i\Delta_a$, $\lambda_m=\kappa_m-i\Delta_m$ and $\lambda=\lambda_a\lambda_m+g^2$. The measured observable considered throughout this work is the cavity amplitude quadrature $\hat X_a=(\hat a+\hat a^\dagger)/\sqrt{2}$ with the mean value
\begin{equation}
\mu(B)\equiv\langle \hat X_a\rangle
=\sqrt{2}\,\mathrm{Re}\,\alpha.
\label{eq:quadrature_mean}
\end{equation}
Throughout this work, we adopt an ideal calibrated quadrature-readout model in which the measured record is expressed in intracavity quadrature units. The analysis therefore characterizes the information contained in the cavity amplitude quadrature $\hat X_a$, assuming that the experimental homodyne chain provides a calibrated linear readout of this quadrature \cite{crescini2021phasemodulated,ebrahimi2021singlequadrature,fan2022optical}. A nonunit detection efficiency $\eta_{\rm det}$ and field-independent added-noise variance $\sigma_{\rm add}^2$ may be included through the mean value of the calibrated quadrature outcomes recorded by the detector $\mu_{\rm meas}$, and their measured variance $\sigma_{\rm meas}^2$
\begin{equation}
\mu_{\rm meas}=\sqrt{\eta_{\rm det}}\,\mu,
\qquad
\sigma_{\rm meas}^2=\eta_{\rm det}\sigma^2+\sigma_{\rm add}^2.
\label{eq:detection_model}
\end{equation}
The ideal results below correspond to $\eta_{\rm det}=1$ and $\sigma_{\rm add}=0$ after calibration. Equation~\eqref{eq:detection_model} makes explicit how detection loss and amplifier noise reduce the experimentally accessible Fisher information without changing its mean--variance decomposition.

The fluctuation operator $\delta\hat a=\hat a-\alpha$ determines the quadrature variance $\sigma^2(B)=\langle[\hat X_a-\mu(B)]^2\rangle$. Solving the steady-state second-moment equations gives
\begin{equation}
\sigma^2
=\left(1-\eta_m\right)n_a^{\mathrm{th}}
+\eta_m n_m^{\mathrm{th}}
+\frac{1}{2},
\label{eq:cavity_variance}
\end{equation}
where the magnon-to-cavity thermal transfer coefficient $\eta_m$ is
\begin{equation}
\eta_m
=\frac{C\kappa_m\kappa}
{(1+C)\kappa^2+\Delta_{am}^2}.
\label{eq:thermal_transfer}
\end{equation}
Here, the cooperativity is $C=g^2/(\kappa_a\kappa_m)$, the total damping rate is $\kappa=\kappa_a+\kappa_m$, and the absolute-detuning difference is $\Delta_{am}\equiv\Delta_a-\Delta_m =\omega_m-\omega_a$. Eq.~\eqref{eq:cavity_variance} expresses the thermal part of the cavity-quadrature variance as a weighted sum of the cavity and magnon thermal occupations. The coefficient $\eta_m$ quantifies the transfer of magnon thermal fluctuations to the cavity and depends on the magnetic field through the cavity--magnon detuning. Because the passive coupling and thermal inputs do not generate phase-sensitive second-order correlations, $\langle\delta\hat a\delta\hat a\rangle=\langle\delta\hat a^{\dagger}\delta\hat a^{\dagger}\rangle=0$. The complete second-moment calculation is given in Appendix~\ref{app:variance}.

The coherent response in Fig.~\ref{fig1}(b) displays the avoided-crossing structure generated by magnon--photon hybridization. Under the symmetric absolute-detuning condition $\Delta_a=\Delta_m\equiv\Delta$, the real parts of the complex poles of $\lambda^{-1}$ in Eq.~\eqref{eq:cavity_amplitude} define the characteristic polariton detunings $\Delta_\pm=\pm\sqrt{g^2-(\kappa_a-\kappa_m)^2/4}$. The cross sections in Fig.~\ref{fig1}(c) consequently develop two resolved branches as the coupling increases, consistent with the established normal-mode structure of cavity-magnon polaritons.\cite{huebl2013high,tabuchi2014hybridizing,zhang2014strongly} These pole frequencies provide the spectral context for magnetic-field transduction, but they do not by themselves identify the operating point of maximum Fisher information.

Because Eqs.~\eqref{eq:cavity_langevin} and \eqref{eq:magnon_langevin} are linear and the input noise is Gaussian, the steady-state amplitude-quadrature outcome $x$ follows the field-conditioned Gaussian distribution
\begin{equation}
p(x|B)
=\frac{1}{\sqrt{2\pi\sigma^2(B)}}
\exp\left[
-\frac{[x-\mu(B)]^2}{2\sigma^2(B)}
\right].
\label{eq:gaussian_likelihood}
\end{equation}
The magnetic field therefore enters the complete measurement statistics through both the coherent mean in Eq.~\eqref{eq:quadrature_mean} and the thermal variance in Eq.~\eqref{eq:cavity_variance}.

\subsection{Fisher Information of the Measured Quadrature}

The classical Fisher information associated with the likelihood in Eq.~\eqref{eq:gaussian_likelihood} is
\begin{equation}
\mathcal{F}(B)
=
\int_{-\infty}^{\infty}
p(x|B)
\left[
\partial_B\ln p(x|B)
\right]^2 dx.
\label{eq:cfi_definition}
\end{equation}
Evaluation of the Gaussian moments gives the exact decomposition\cite{nielsen2025variational}
\begin{equation}
\mathcal{F}(B)
=\frac{[\partial_B\mu(B)]^2}{\sigma^2(B)} + \frac{[\partial_B\sigma^2(B)]^2}{2\sigma^4(B)}.
\label{eq:cfi_decomposition}
\end{equation}
For $N$ statistically independent quadrature samples, every locally unbiased estimator obeys the Cram\'er--Rao inequality
\begin{equation} 
\delta B\equiv\sqrt{\operatorname{Var}(\widehat B)}\geq\delta B_{\rm CR},\qquad \delta B_{\rm CR}\equiv\frac{1}{\sqrt{N\mathcal{F}(B)}}.
\label{eq:cramer_rao}
\end{equation}
{  The Fisher information therefore sets a lower bound on the field-estimation uncertainty of the stated readout protocol. For a continuous homodyne record, we use the effective independent-sample description detailed in Appendix~\ref{app:sensitivity}. The time-normalized Cram\'er--Rao sensitivity bound is $\mathcal{S}\equiv\delta B_{\rm CR}(\tau)\sqrt{\tau}$, where $\tau$ is the integration time. The measured field uncertainty consequently satisfies $\delta B(\tau)\sqrt{\tau}\geq\mathcal{S}$.}

The coherent-displacement contribution $\mathcal{F}_{\rm coh}$ is determined by the field susceptibility of the quadrature mean, whereas the thermal-fluctuation contribution $\mathcal{F}_{\rm th}$ is determined by the field susceptibility of the quadrature variance. The two terms are statistically distinct contributions to the same measured distribution rather than mutually exclusive measurement protocols. The Gaussian integration leading to Eq.~\eqref{eq:cfi_decomposition} is provided in Appendix~\ref{app:cfi}.

To express the coherent susceptibility explicitly, we introduce
\begin{equation}
\begin{aligned}
&A=\kappa_a\kappa_m-\Delta_a\Delta_m+g^2, \\
&D=\kappa_a\Delta_m+\kappa_m\Delta_a,
\label{eq:A_D_definitions}
\end{aligned}
\end{equation}
Differentiating Eq.~\eqref{eq:cavity_amplitude} with respect to the magnon detuning gives
\begin{equation}
\partial_B\mu=
-\gamma\frac{\partial\mu}{\partial\Delta_m}
= -\gamma
\frac{
2\sqrt{2}\,\Omega g^2 A D
}{\left(A^2+D^2\right)^2}.
\label{eq:mean_field_derivative}
\end{equation}
The factor $AD$ connects the field sensitivity of the selected amplitude quadrature to the dispersive structure of the coupled susceptibility. In particular, the coherent Fisher information depends on the derivative of the measured quadrature rather than on the magnitude or peak position of the coherent spectrum alone.

The variance contains two independent sources of field dependence. Defining the thermal occupation imbalance
\begin{equation}
\Delta n^{\mathrm{th}}
= n_m^{\mathrm{th}}-n_a^{\mathrm{th}}
\label{eq:thermal_imbalance}
\end{equation}
and the local magnon Bose-occupation frequency-susceptibility scale
\begin{equation}
\xi_m
=n_m^{\mathrm{th}}
\left(n_m^{\mathrm{th}}+1\right)
\frac{\hbar}{k_BT},
\label{eq:thermal_susceptibility}
\end{equation}
the exact field derivative of Eq.~\eqref{eq:cavity_variance} becomes
\begin{equation}
\partial_B\sigma^2
=-\gamma
\left[
\Delta n^{\mathrm{th}}
\frac{\partial\eta_m}{\partial\Delta_m}
+\eta_m\xi_m
\right].
\label{eq:variance_field_derivative}
\end{equation}
The transfer-coefficient derivative appearing here is
\begin{equation}
\frac{\partial\eta_m}{\partial\Delta_m}
=\frac{
2C\kappa_m\kappa\Delta_{am}}{\left[(1+C)\kappa^2+\Delta_{am}^2\right]^2}.
\label{eq:transfer_derivative}
\end{equation}
The first term in Eq.~\eqref{eq:variance_field_derivative} describes the field-dependent redistribution of thermal fluctuations between the two modes in the presence of an occupation imbalance. The second describes the direct change of the magnon Bose occupation as its physical frequency shifts with the field. Consequently, matching the two thermal occupations removes the first mechanism but not the second: at finite temperature, $\Delta n^{\mathrm{th}}=0$ does not imply $\mathcal{F}_{\rm th}=0$.

Equations~\eqref{eq:mean_field_derivative} and \eqref{eq:variance_field_derivative} identify coherent displacement and thermal variance as simultaneous carriers of magnetic-field information in the same cavity-quadrature statistics. The next step is to determine whether their distinct susceptibilities favor the same detuning and coupling conditions.

\section{Coherent-Response and Thermal-Fluctuation Information}

Having identified the two contributions in Eq.~\eqref{eq:cfi_decomposition}, we now compare their dependence on coupling and detuning under the same device constraints. The cavity damping rate $\kappa_a$ and magnon damping rate $\kappa_m$ are treated as fixed parameters, while the cooperativity $C$ is varied through the coupling strength $g$. The drive frequency and static magnetic field provide control over the cavity and magnon detunings. For the analytical optimization of the coherent morphology, the amplitude-quadrature variance $\sigma^2$ is treated as constant because its variation over the parameter range considered is weak compared with the resonant change of the coherent susceptibility. The full field dependence of $\sigma^2$ is retained when evaluating the thermal contribution and the total Fisher information.

\subsection{Direct Comparison of the Two Information Contributions}

\begin{figure}[htbp]
\centering
\includegraphics[width=\linewidth]{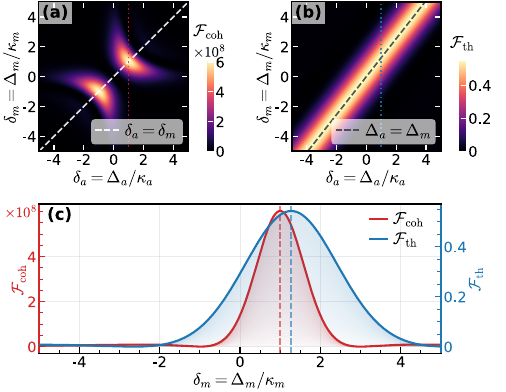}
\caption{Comparison of the two contributions to the measured-quadrature Fisher information. (a) The coherent-displacement contribution $\mathcal{F}_{\rm coh}$ at $C=2$ is concentrated near $\delta_a=\delta_m$. (b) The thermal-fluctuation contribution $\mathcal{F}_{\rm th}$ forms a dominant ridge near $\Delta_a=\Delta_m$, corresponding to $\delta_m=(\kappa_a/\kappa_m)\delta_a$ in normalized coordinates. (c) Cuts at $C=2$ and $\delta_a=1$ show the offset between the coherent maximum at $\delta_m=1$ and the thermal maximum near $\delta_m\simeq1.27$. The plotted quantities are the corresponding Fisher-information contributions, with the same normalization used on the color bars and axes. We take $T=300$ K and the other parameters are the same as those in Fig.~\ref{fig1}.}
\label{fig:competition}
\end{figure}
The exact decomposition in Eq.~\eqref{eq:cfi_decomposition} already suggests two different operating principles because $\partial_B\mu$ is governed by the coherent coupled-mode susceptibility, whereas $\partial_B\sigma^2$ depends on the absolute cavity--magnon frequency difference and the Bose-occupation susceptibility. This distinction is visible before any reduced optimization is introduced. In Fig.~\ref{fig:competition}(a), the coherent contribution is concentrated near the linewidth-normalized matching line $\delta_a=\delta_m$. The thermal contribution in Fig.~\ref{fig:competition}(b) instead forms a dominant ridge near absolute frequency matching, $\Delta_a=\Delta_m$. In normalized coordinates, the latter condition is $\delta_m=(\kappa_a/\kappa_m)\delta_a$ and therefore differs from $\delta_a=\delta_m$ whenever $\kappa_a\neq\kappa_m$.

The one-dimensional cuts in Fig.~\ref{fig:competition}(c) make the experimental consequence explicit. At $C=2$ and $\delta_a=1$, the coherent contribution peaks at $\delta_m=1$, whereas the thermal contribution peaks near $\delta_m=\kappa_a/\kappa_m\simeq1.27$ for the damping rates used here. The two profiles overlap, so both contributions can remain appreciable in a common configuration, but their distinct maxima provide a direct operating choice. The following subsections explain and analytically optimize these two structures.

\subsection{Coherent-Displacement Contribution}

The decomposition in Eq.~\eqref{eq:cfi_decomposition} shows that the coherent contribution is set by the magnetic-field derivative of the cavity-quadrature mean. Using the linewidth-normalized detunings defined in Eq.~\eqref{eq:normalized_detunings}, Eq.~\eqref{eq:mean_field_derivative} gives
\begin{equation}
\mathcal{F}_{\rm coh}
=\gamma^2
\left(
\frac{2\Omega^2}
{\sigma^2\kappa_a^2\kappa_m^2}
\right)
f_{\rm coh}(C,\delta_a,\delta_m),
\label{eq:coherent_cfi_factorization}
\end{equation}
where the dimensionless coherent morphology is
\begin{equation}
f_{\rm coh}(C,\delta_a,\delta_m)
=\frac{
4C^2(\delta_a+\delta_m)^2
(1+C-\delta_a\delta_m)^2
}{\left[
(1+C-\delta_a\delta_m)^2
+(\delta_a+\delta_m)^2
\right]^4}.
\label{eq:coherent_morphology}
\end{equation}
The prefactor in Eq.~\eqref{eq:coherent_cfi_factorization} sets the overall information scale through the gyromagnetic ratio, drive amplitude, damping rates, and quadrature variance. The function $f_{\rm coh}$ determines how this information is distributed over coupling and detuning.

\begin{figure}[htbp]
\centering
\includegraphics[width=\linewidth]{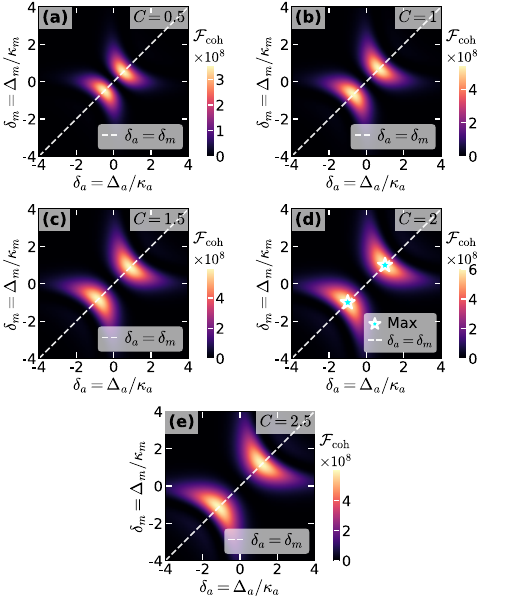}
\caption{Coherent Fisher information in the normalized-detuning plane $(\delta_m,\delta_a)$ for cooperativities ranging from $C=0.5$ to $C=2.5$. The high-information regions follow the two hybridized response branches as the coupling increases. The white dashed line marks the normalized-detuning condition $\delta_a=\delta_m$, along which the largest values occur. The other parameters are the same as those in Fig.~\ref{fig:competition}.}
\label{fig:coh_fisher_2d}
\end{figure}

The coherent-information shown in Fig.~\ref{fig:coh_fisher_2d} follows the mode-splitting response of the coupled cavity--magnon modes. At smaller cooperativity, the high-information regions remain close to the unsplit resonance. As $C$ increases, they separate into two branches associated with the hybridized susceptibility. Their largest values lie on the normalized-detuning line $\delta_a=\delta_m$, which differs from the absolute-detuning condition $\Delta_a=\Delta_m$ when the two damping rates are unequal. The contour landscape identifies the relevant region of parameter space, while the location of the maximum follows from the stationary conditions below.

To determine the global maximum of $f_{\rm coh}$, we require its partial derivatives with respect to the normalized detunings to vanish simultaneously
\begin{equation}
\frac{\partial f_{\rm coh}}
{\partial\delta_a}=\frac{\partial f_{\rm coh}}
{\partial\delta_m}=0.
\label{eq:coherent_detuning_stationary_conditions}
\end{equation}
Solving these conditions for the nonzero extrema gives the symmetry relation
\begin{equation}
\delta_a
=\delta_m
\equiv \delta_{\rm sym}.
\label{eq:coherent_symmetry_line}
\end{equation}
The detailed reduction of the two stationary equations is given in Appendix~\ref{app:coherent_optimization}. Substituting Eq.~\eqref{eq:coherent_symmetry_line} into Eq.~\eqref{eq:coherent_morphology} reduces the optimization to
\begin{equation}
f_{\rm coh}(C,\delta_{\rm sym})
=\frac{
16C^2\delta_{\rm sym}^2
(1+C-\delta_{\rm sym}^2)^2
}{\left[
(1+C-\delta_{\rm sym}^2)^2
+4\delta_{\rm sym}^2
\right]^4}.
\label{eq:coherent_reduced_morphology}
\end{equation}

\begin{figure}[htbp]
\centering
\includegraphics[width=\linewidth]{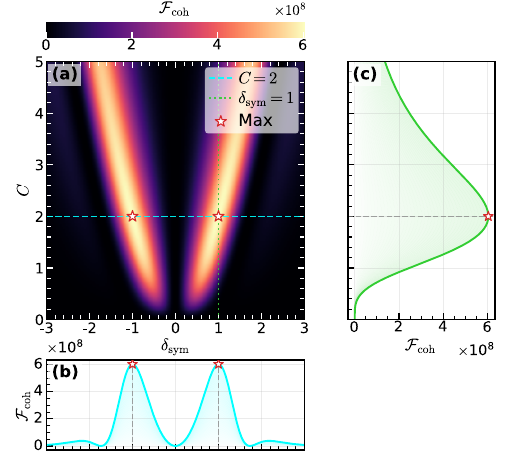}
\caption{Optimization of the coherent Fisher-information morphology in the reduced parameter space $(C,\delta_{\rm sym})$. (a) The two-dimensional landscape contains two maxima at $(C,\delta_{\rm sym})=(2,\pm1)$. (b) The detuning cut at $C=2$ shows the two maxima and the zero at $\delta_{\rm sym}=0$. (c) The coupling cut at $\delta_{\rm sym}=1$ reaches its maximum at $C=2$ and decreases for larger $C$. The other parameters are the same as those in Fig.~\ref{fig:competition}.}
\label{fig:coh_fisher_optimization}
\end{figure}

The remaining stationary conditions are
\begin{equation}
\frac{\partial f_{\rm coh}(C,\delta_{\rm sym})}{\partial C}=\frac{\partial f_{\rm coh}(C,\delta_{\rm sym})}{\partial\delta_{\rm sym}}=0,
\label{eq:coherent_reduced_stationary_conditions}
\end{equation}
Solving these equations and comparing the stationary values with the boundaries of the physical parameter domain gives
\begin{equation}
\begin{aligned}
C_{\rm opt}
&=2,
\\
\delta_{a,\rm opt}
&=\delta_{m,\rm opt}=
\pm1.
\end{aligned}
\label{eq:coherent_global_optimum}
\end{equation}
Thus, the coherent displacement of the chosen amplitude quadrature carries the most field information at finite coupling and finite normalized detuning. When the two damping rates are fixed and $g$ is the tunable parameter, the optimal cooperativity corresponds to
\begin{equation}
g_{\rm opt}
=
\sqrt{2\kappa_a\kappa_m}.
\label{eq:optimal_coupling}
\end{equation}
{  The normalized-detuning condition $\delta_a=\delta_m$ reflects the role of damping in shaping the driven response. Each mode responds to detuning on the frequency scale set by its own linewidth, which controls the response width and phase variation. Matching the normalized detunings therefore places the two modes at corresponding positions within their response profiles, providing a physical interpretation of the optimum for the selected amplitude quadrature.

The two localized maxima predicted by Eq.~\eqref{eq:coherent_global_optimum} appear in the $(C,\delta_{\rm sym})$ landscape of Fig.~\ref{fig:coh_fisher_optimization}(a). The detuning cut at $C=2$ in Fig.~\ref{fig:coh_fisher_optimization}(b) passes through both maxima and through a zero at $\delta_{\rm sym}=0$. The coupling cut at $\delta_{\rm sym}=1$ in Fig.~\ref{fig:coh_fisher_optimization}(c) rises to its maximum at $C=2$ and decreases at larger cooperativity. These features follow directly from Eq.~\eqref{eq:coherent_reduced_morphology}, which gives $f_{\rm coh}(C,0)=0$, $f_{\rm coh}(C,1)=16C^4 / (C^2+4)^4$. The zero at $\delta_{\rm sym}=0$ results from the vanishing field derivative of the chosen amplitude-quadrature mean along this symmetric path and is therefore a zero of $\mathcal{F}_{\rm coh}$ for this measurement configuration. The decrease beyond \(C=2\) follows directly from the coupling dependence of the selected amplitude-quadrature susceptibility.}

\subsection{Thermal-Fluctuation Contribution}

The coherent-displacement contribution reaches its maximum on the normalized-detuning line $\delta_a=\delta_m$. The thermal-variance contribution follows a different parameter dependence because Eq.~\eqref{eq:variance_field_derivative} contains both the magnon-to-cavity transfer coefficient and the magnon thermal occupation. The term proportional to $\Delta n^{\mathrm{th}}\partial_{\Delta_m}\eta_m$ describes the change in thermal-fluctuation transfer as the magnon frequency shifts, whereas the term $\eta_m\xi_m$ describes the direct field dependence of the magnon Bose occupation. Both mechanisms contribute to the variance of the same cavity quadrature.

The origin of the thermal ridge can be obtained from the exact variance derivative. We define the dimensionless absolute-detuning difference
\begin{equation}
x_{\rm eff}=\frac{\Delta_{am}}{\kappa}=\frac{\omega_m-\omega_a}{\kappa}.
\label{eq:effective_detuning}
\end{equation}
Using Eqs.~\eqref{eq:thermal_transfer}, \eqref{eq:variance_field_derivative}, and \eqref{eq:effective_detuning}, the thermal Fisher information can be factored as
\begin{equation}
\mathcal{F}_{\rm th}=\gamma^2\left(\frac{\kappa_m^2}{2\sigma^4\kappa^4}\right)f_{\rm th}(C,x_{\rm eff}),
\label{eq:incoherent_cfi_factorization}
\end{equation}
where
\begin{equation}
f_{\rm th}(C,x_{\rm eff})=\frac{C^2\left[2x_{\rm eff}\Delta n^{\mathrm{th}}+\xi_m\kappa\left(1+C+x_{\rm eff}^2\right)\right]^2}{\left(1+C+x_{\rm eff}^2\right)^4}.
\label{eq:exact_incoherent_morphology}
\end{equation}
Eq.~\eqref{eq:exact_incoherent_morphology} retains both sources of thermal-variance susceptibility. In particular, the term containing $\xi_m$ remains finite at nonzero temperature even when the thermal occupation imbalance vanishes. At frequency matching, the common-temperature occupations satisfy $n_m^{\mathrm{th}}=n_a^{\mathrm{th}}$, but a small field variation still changes $n_m^{\mathrm{th}}$. The variance can therefore retain local magnetic-field information at the point where the two occupations are equal.

To obtain the detuning dependence analytically, we expand the magnon Bose occupation around $\omega_m=\omega_a$. With the cavity Bose-occupation frequency-susceptibility scale $\xi_a=n_a^{\mathrm{th}}(n_a^{\mathrm{th}}+1)\hbar/(k_BT)$, the first-order expansion gives
\begin{equation}
\begin{aligned}
\Delta n^{\mathrm{th}}
&\simeq -\xi_a\kappa x_{\rm eff},
\\
\xi_m
&\simeq \xi_a.
\end{aligned}
\label{eq:local_thermal_expansion}
\end{equation}
This expansion requires $|\omega_m-\omega_a|=|\kappa x_{\rm eff}|\ll\omega_a$. The detunings considered here are on the megahertz scale, whereas the cavity and magnon carrier frequencies are on the gigahertz scale, so the expansion is suitable for the parameter range of the figures.

Substituting Eq.~\eqref{eq:local_thermal_expansion} into Eq.~\eqref{eq:exact_incoherent_morphology} gives
\begin{equation}
\mathcal{F}_{\rm th}
\simeq \gamma^2
\left( \frac{\xi_a^2\kappa_m^2}
{2\sigma^4\kappa^2}
\right)
\widetilde f_{\rm th}(C,x_{\rm eff}),
\label{eq:reduced_incoherent_cfi}
\end{equation}
with the reduced thermal morphology
\begin{equation}
\widetilde f_{\rm th}(C,x_{\rm eff})
=\frac{C^2\left(1+C-x_{\rm eff}^2
\right)^2}{\left(1+C+x_{\rm eff}^2
\right)^4}.
\label{eq:reduced_incoherent_morphology}
\end{equation}
Equation~\eqref{eq:reduced_incoherent_cfi} separates the temperature-dependent and damping-dependent scale from the dimensionless dependence on coupling and relative detuning. Differentiating Eq.~\eqref{eq:reduced_incoherent_morphology} gives
\begin{equation}
\frac{
\partial\widetilde f_{\rm th}
}{\partial x_{\rm eff}}
=-\frac{4C^2x_{\rm eff}
\left(1+C-x_{\rm eff}^2\right)
\left[3(1+C)-x_{\rm eff}^2
\right]}{\left(1+C+x_{\rm eff}^2
\right)^5}.
\label{eq:incoherent_detuning_derivative}
\end{equation}
For the reduced morphology obtained from the near-resonance expansion, the central stationary point in Eq.~\eqref{eq:incoherent_detuning_derivative} gives the global maximum,
\begin{equation}
x_{\rm eff}^{\rm opt}=0
\quad \Longleftrightarrow \quad
\Delta_a=\Delta_m.
\label{eq:incoherent_optimal_detuning}
\end{equation}
{  The additional stationary points satisfy $x_{\rm eff}^2=1+C$ and $x_{\rm eff}^2=3(1+C)$ (see Appendix~\ref{app:incoherent_optimization} for details). The former are zeros of the reduced morphology, whereas the latter are weak side maxima with a height equal to $1/64$ of the central maximum. The profile is therefore dominated by the central ridge at $\Delta_a=\Delta_m$.} {  This thermal optimum follows the actual frequency matching of the two modes. At the matching point, thermal fluctuations are exchanged most efficiently between the cavity and magnon modes. The field-induced change in the magnon thermal occupation is therefore transferred to the cavity variance with the largest weight. This is different from the coherent channel, where damping shapes the driven response and the optimal detunings are matched relative to individual linewidths.

Along the optimal-detuning line, Eq.~\eqref{eq:reduced_incoherent_morphology} reduces to
\begin{equation}
\widetilde f_{\rm th}^{\rm opt}(C)
=\widetilde f_{\rm th}(C,0)
=\frac{C^2}{(1+C)^2}.
\label{eq:optimized_incoherent_morphology}
\end{equation}
Within the adopted local dissipative model, the optimized thermal morphology increases with $C$ and approaches the finite limit $\lim_{C\rightarrow\infty}\widetilde f_{\rm th}^{\rm opt}=1$. This asymptotic result applies only to the reduced morphology.}

\begin{figure}[htbp]
\centering
\includegraphics[width=\linewidth]{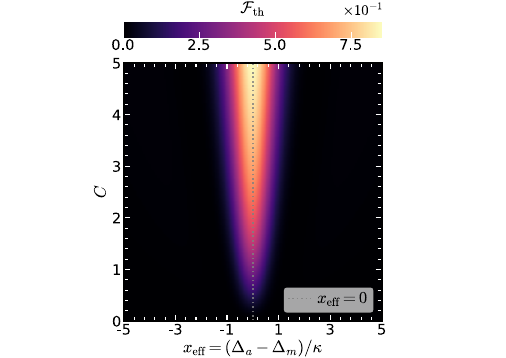}
\caption{Thermal Fisher-information structure in the reduced parameter space $(x_{\rm eff},C)$. The dominant ridge lies at $x_{\rm eff}=0$, corresponding to the absolute-detuning condition $\Delta_a=\Delta_m$. Along this ridge, the reduced morphology is $C^2/(1+C)^2$ and approaches a finite limiting value as $C$ increases. Weak side maxima are present but remain much smaller than the central ridge. The other parameters are the same as those in Fig.~\ref{fig:competition}.}
\label{fig:thermal_dispersion}
\end{figure}

Fig.~\ref{fig:thermal_dispersion} displays the resulting structure in the $(x_{\rm eff},C)$ plane. The dominant ridge remains centered at $x_{\rm eff}=0$, as required by Eq.~\eqref{eq:incoherent_optimal_detuning}. Its height follows Eq.~\eqref{eq:optimized_incoherent_morphology}, increasing rapidly at smaller $C$ and changing more slowly as it approaches its finite limiting value. The much weaker side maxima do not shift the global optimum away from absolute-detuning matching.

The thermal contribution has a clear zero-information boundary. At zero temperature, both reservoirs are in their vacuum states and the quadrature variance loses its magnetic-field dependence $\mathcal{F}_{\rm th}=0$. This boundary differs from occupation matching at finite temperature, where the local Bose-occupation susceptibility remains nonzero. The two operating principles are summarized in Table~\ref{tab:operating_rules}. The coherent-displacement contribution follows linewidth-normalized detuning matching, whereas the thermal-fluctuation contribution follows absolute frequency matching within the near-resonance expansion. Their combined contribution at a common high-information point is examined next.

\begin{table*}[htbp]
\caption{Measurement-specific operating rules for the two Fisher-information contributions.}
\label{tab:operating_rules}
\begin{ruledtabular}
\begin{tabular}{lll}
Contribution & Preferred condition & Physical origin \\
\hline
$\mathcal{F}_{\rm coh}$ & $C=2$, $\delta_a=\delta_m=\pm1$ & Mean susceptibility \\
$\mathcal{F}_{\rm th}$ & $\Delta_a=\Delta_m$ & Variance susceptibility
\end{tabular}
\end{ruledtabular}
\end{table*}

\section{Total Fisher Information and Sensitivity Scaling}

The preceding optimization identifies different detuning conditions for the coherent-displacement and thermal-fluctuation contributions. To evaluate the information available from both contributions in the same device configuration, we fix the cooperativity at $C=2$ and the normalized detunings at $\delta_a=\delta_m=1$. This point maximizes the coherent morphology and lies close to the central thermal ridge for the damping rates considered here, but it does not independently maximize both contributions. We therefore refer to it as a common near-optimal operating configuration. Following the exact decomposition in Eq.~\eqref{eq:cfi_decomposition}, we denote the total cavity amplitude-quadrature Fisher information by $\mathcal{F}_{\rm total}=\mathcal{F}_{\rm coh}+\mathcal{F}_{\rm th}$. We examine how temperature and coherent driving determine this total information and the sensitivity improvement supplied by the thermal contribution.

{Using the effective independent-sample description, the corresponding Cram\'er--Rao field uncertainty for an effective integration time $\tau$ is (see details in Appendix \ref{app:sensitivity})
\begin{equation} 
\delta B_{\rm CR}(\tau)
=\sqrt{\frac{\tau_{\rm corr}}{\tau\,\mathcal{F}_{\rm total}}},
\delta B_{\rm CR}\sqrt{\tau}
=\sqrt{\frac{\tau_{\rm corr}}{\mathcal{F}_{\rm total}}}=\mathcal{S}_{\rm total},
\label{eq:sensitivity_rate}
\end{equation}
where $\tau_{\rm corr}$ is the effective correlation time of the calibrated output record. Equation~\eqref{eq:sensitivity_rate} converts every information landscape below into a magnetometric sensitivity once the experimental bandwidth is specified. It also makes clear that the single-sample Fisher information alone should not be interpreted as a sensitivity per $\sqrt{\mathrm{Hz}}$ without a temporal measurement model.}

The information scales entering Eq.~\eqref{eq:sensitivity_rate} follow from the factorizations in Eq.~\eqref{eq:coherent_cfi_factorization} and Eq.~\eqref{eq:reduced_incoherent_cfi}. At the common operating configuration, the coherent scale is proportional to $2\Omega^2/(\sigma^2\kappa_a^2\kappa_m^2)$, whereas the thermal scale within the local thermal expansion is proportional to $\xi_a^2\kappa_m^2/(2\sigma^4\kappa^2)$. Dissipation therefore enters the two contributions differently: the coherent scale contains inverse powers of both damping rates, while the thermal scale depends on the linewidth ratio $\kappa_m/\kappa$ together with the thermal susceptibility and quadrature variance. These damping rates characterize the dissipative conditions of the chosen device configuration. With the device parameters fixed, temperature and coherent driving control the information scales through $\sigma^2$, $\xi_a$, and $\Omega$.

Temperature changes the two contributions through different parts of the quadrature distribution. The coherent mean is independent of temperature when the drive and device parameters are fixed, but the thermal growth of the variance $\sigma^2$ reduces the coherent contribution through the denominator of $\mathcal{F}_{\rm coh}$. At the same time, the field dependence of the magnon Bose occupation generates the variance derivative in Eq.~\eqref{eq:variance_field_derivative} and thereby supports $\mathcal{F}_{\rm th}$. In the zero-temperature limit, the thermal occupations and their frequency susceptibilities vanish, so $\mathcal{F}_{\rm th}$ approaches zero. At finite temperature, thermal broadening lowers the information carried by a fixed coherent displacement, while the field-dependent thermal variance provides an additional contribution. {  The total Fisher information therefore reflects both effects.}

\begin{figure}[htbp]
\centering
\includegraphics[width=\linewidth]{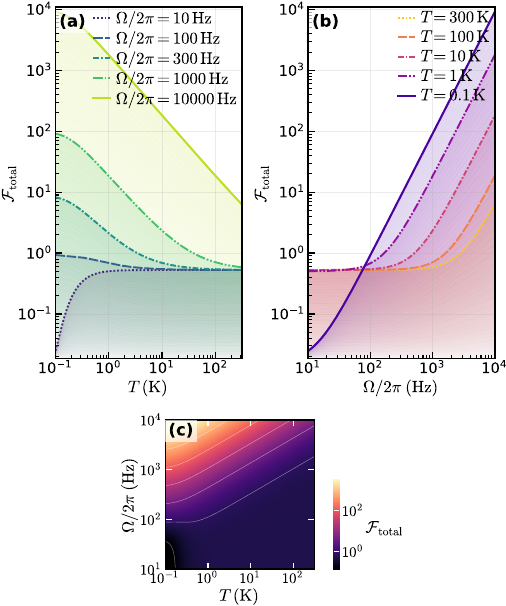}
\caption{Temperature and drive dependence of the total cavity amplitude-quadrature Fisher information at the common near-optimal configuration $C=2$ and $\delta_a=\delta_m=1$. (a) Temperature dependence of $\mathcal{F}_{\rm total}$ for several coherent drive amplitudes. Thermal broadening reduces the coherent contribution, while the field-dependent thermal variance supplies the thermal contribution at finite temperature. (b) Drive dependence of $\mathcal{F}_{\rm total}$ for several temperatures. A finite thermal-information baseline remains at $\Omega=0$ when $T>0$, and the coherent contribution increases with $\Omega^2$ within the model-valid drive range. (c) Map of $\log_{10}\mathcal{F}_{\rm total}$ in the $(T,\Omega)$ plane. The contours mark constant total Fisher information and show the combined effect of thermal broadening and coherent driving. Parameters not varied in the panels are the same as those in Fig.~\ref{fig1}.}
\label{fig:temperature_drive}
\end{figure}
This balance appears in Fig.~\ref{fig:temperature_drive}(a), which shows $\mathcal{F}_{\rm total}$ as a function of temperature for several drive amplitudes. At weaker drive, the coherent contribution is small and the finite-temperature variation is set mainly by the thermal susceptibility of the quadrature variance. At stronger drive, the low-temperature response is larger because the coherent displacement is measured against a smaller variance. Increasing temperature then reduces this coherent information through thermal broadening, while the thermal contribution supplies a finite thermal component. The resulting curves describe the total information available at each temperature, with the relative contributions of the coherent and thermal channels varying across the temperature range.

The drive dependence is more direct. According to Eq.~\eqref{eq:coherent_cfi_factorization}, $\mathcal{F}_{\rm coh}$ is proportional to $\Omega^2$ when the remaining parameters are fixed. The covariance of the passive thermal state does not depend on the coherent displacement in the adopted model, so $\mathcal{F}_{\rm th}$ remains independent of $\Omega$. The total Fisher information in Fig.~\ref{fig:temperature_drive}(b) therefore approaches a drive-independent thermal baseline as $\Omega$ tends to zero and increases as the coherent-displacement contribution is added. At finite temperature, the zero-drive limit gives the important boundary
\begin{equation}
\Omega=0,\quad T>0 \quad\Longrightarrow\quad \mathcal{F}_{\rm total}=\mathcal{F}_{\rm th}>0. \label{eq:zero_drive_thermal_information}
\end{equation}
Eq.~\eqref{eq:zero_drive_thermal_information} shows that the field-dependent thermal variance retains magnetic-field information even without a coherent cavity displacement. It does not identify zero drive as the best sensing condition. Increasing $\Omega$ adds coherent information as long as the drive remains within the regime where neglected nonlinearities are small.

The joint dependence on temperature and drive is summarized by the map of $\log_{10}\mathcal{F}_{\rm total}$ in Fig.~\ref{fig:temperature_drive}(c). Within the plotted range, the largest total Fisher information occurs at lower temperature and stronger drive, where the coherent response is measured against reduced thermal variance. Toward weaker drive, the total information approaches the finite-temperature baseline supplied by the thermal contribution. The constant-information contours also show that a larger drive is generally required to maintain the same total Fisher information as thermal broadening increases. {  To quantify how much the thermal contribution improves the corresponding field uncertainty, we next compare the two contributions at the same operating conditions.}

\begin{figure}[htbp]
\centering
\includegraphics[width=\linewidth]{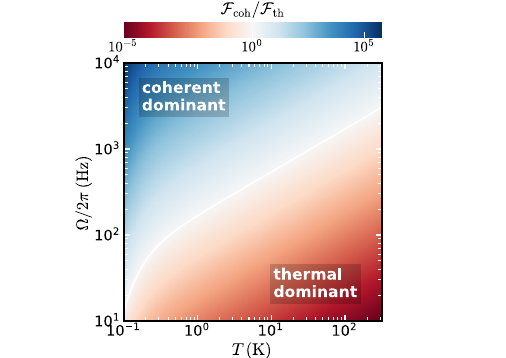} 
\caption{Relative contributions of the coherent-displacement and thermal-variance Fisher information as functions of temperature and coherent drive. The color scale shows the ratio $\mathcal{F}_{\rm coh}/\mathcal{F}_{\rm th}$ at the common operating configuration $C=2$ and $\delta_a=\delta_m=1$. Blue regions correspond to $\mathcal{F}_{\rm coh}>\mathcal{F}_{\rm th}$, where the magnetic-field information is dominated by the coherent displacement, while red regions correspond to $\mathcal{F}_{\rm th}>\mathcal{F}_{\rm coh}$, where the field-dependent thermal variance provides the dominant contribution. The white contour marks $\mathcal{F}_{\rm coh}=\mathcal{F}_{\rm th}$. The map shows that strong coherent driving favors displacement-based sensing, whereas the thermal contribution becomes comparatively important at sufficiently weak drive and finite temperature. Parameters are the same as those in Fig.~\ref{fig:temperature_drive}.}
\label{fig:drive_regime}
\end{figure}

Figure~\ref{fig:drive_regime} shows the ratio $r=\mathcal{F}_{\rm coh}/\mathcal{F}_{\rm th}$ in the $(T,\Omega)$ plane. The white contour marks $r=1$, where the coherent displacement and thermal variance contribute equally. The blue and red regions correspond to coherent and thermal dominance, respectively. Since $\mathcal{F}_{\rm coh}\propto\Omega^2$ and $\mathcal{F}_{\rm th}$ is independent of $\Omega$, reducing the drive increases the relative weight of the thermal contribution. The map therefore identifies the weak-drive conditions under which field-dependent thermal fluctuations become the principal source of information in the measured cavity quadrature.

The ratio $r$ directly determines the improvement in the Cram\'er--Rao sensitivity bound. For the same effective sampling rate, denote the time-normalized bounds obtained from the coherent contribution and total information by $\mathcal{S}_{\rm coh}=\sqrt{\tau_{\rm corr}/\mathcal{F}_{\rm coh}}$ and $\mathcal{S}_{\rm total}=\sqrt{\tau_{\rm corr}/\mathcal{F}_{\rm total}}$, respectively. Their ratio is
\begin{equation}
\frac{\mathcal{S}_{\rm total}}{\mathcal{S}_{\rm coh}}=\sqrt{\frac{\mathcal{F}_{\rm coh}}{\mathcal{F}_{\rm coh}+\mathcal{F}_{\rm th}}}=\sqrt{\frac{r}{1+r}}.
\label{eq:sensitivity_gain_ratio}
\end{equation}
Thus, the white contour in Fig.~\ref{fig:drive_regime} corresponds to a factor of two increase in Fisher information and a $29.3\%$ reduction of the standard-deviation bound. In the thermal-dominated region, the reduction is larger. This comparison quantifies the precision gained by including the field-dependent variance in the same quadrature measurement.

For a representative point at $T=100\,\mathrm{K}$ and $\Omega/2\pi=10^3\,\mathrm{Hz}$, the two contributions are $\mathcal{F}_{\rm coh}\simeq0.181\,\mathrm{T}^{-2}$ and $\mathcal{F}_{\rm th}\simeq0.528\,\mathrm{T}^{-2}$. Their sum is $\mathcal{F}_{\rm total}\simeq0.709\,\mathrm{T}^{-2}$, giving $r\simeq0.343$ and $\mathcal{S}_{\rm total}/\mathcal{S}_{\rm coh}\simeq0.505$. Including the thermal contribution therefore lowers the standard-deviation bound by approximately $49.5\%$.  This point illustrates how the thermal variance contributes substantially to the available information at weak coherent drive. Together, Figs.~\ref{fig:temperature_drive} and \ref{fig:drive_regime} connect the total information set by temperature and driving to the precision gained from the field-dependent thermal fluctuations.

\section{Conclusion and Discussion}
We have developed a measurement-specific Fisher-information framework for longitudinal magnetic-field sensing with a cavity--magnon system. The central result is the exact separation of the Gaussian quadrature Fisher information into a coherent-displacement contribution $\mathcal{F}_{\rm coh}$, governed by the field dependence of the measured mean, and a thermal-fluctuation contribution $\mathcal{F}_{\rm th}$, governed by the field dependence of the measured variance. These statistically distinct contributions belong to the same calibrated quadrature record rather than to separate measurement protocols. Thermal fluctuations therefore play a dual metrological role: they reduce coherent-displacement information by broadening the measured distribution, while their field-dependent statistics provide an additional source of magnetic-field information.

The two contributions lead to distinct operating principles. Under the constant-variance optimization of the selected amplitude quadrature, the dimensionless coherent morphology is globally maximized at cooperativity $C=2$ and linewidth-normalized detunings $\delta_a=\delta_m=\pm1$, corresponding to $g=\sqrt{2\kappa_a\kappa_m}$ for fixed damping rates. Within the near-resonance expansion, the reduced thermal-fluctuation morphology is globally maximized at $x_{\rm eff}=0$, equivalent to absolute cavity--magnon frequency matching, $\Delta_a=\Delta_m$. Equal cavity and magnon thermal occupations at this operating point do not eliminate the variance information because the local frequency susceptibility of the magnon thermal occupation remains finite. Consequently, at finite temperature $\mathcal{F}_{\rm th}$ remains nonzero even when the coherent drive is removed, giving $\mathcal{F}_{\rm total}=\mathcal{F}_{\rm th}>0$ at $\Omega=0$, whereas the thermal-fluctuation contribution vanishes in the zero-temperature limit. This result demonstrates that thermal fluctuations can act as an operational information carrier rather than solely as a source of measurement noise.

The relative behavior of the coherent-drive and thermal-fluctuation contributions, illustrated in Fig.~\ref{fig:drive_regime} highlights the role of thermal fluctuations as carriers of magnetic-field information. When the coherent response is weak, the field-dependent variance can provide an additional sensing channel, provided that detection noise is sufficiently controlled. 

The coupling--detuning and temperature--drive maps consequently provide practical guidance for selecting an operating point according to whether coherent response, thermal fluctuations, or both are to be exploited. The results apply to a calibrated amplitude-quadrature measurement with passive cavity--magnon coupling, Gaussian local thermal dissipation, slowly varying fields, and drive strengths within the linear-response regime specified by Eq.~\eqref{eq:linear_response_condition}. Detection inefficiency and added amplifier noise can be incorporated through Eq.~\eqref{eq:detection_model}, while conversion of the single-sample Fisher information into a sensitivity per $\sqrt{\mathrm{Hz}}$ requires the output correlation time introduced in Eq.~\eqref{eq:sensitivity_rate}. Future work may optimize the homodyne angle, evaluate the Fisher-information rate of the complete continuous measurement record, compare the quadrature-accessible information with the quantum Fisher information, and examine the robustness of the thermal-information contribution under nonlinear and global dissipative models.

\begin{acknowledgments}
We acknowledge financial support from NSF grants PHY-2316878, PHY-2514953.
\end{acknowledgments}

\section*{Data Availability Statement}
Data sharing is not applicable to this article because no new experimental data were created or analyzed in this theoretical study.

\appendix

\section{Steady-State Fluctuation Moments and Cavity Variance}
\label{app:variance}

We derive the cavity-quadrature variance used in Eq.~\eqref{eq:cavity_variance} from the steady-state second moments of the coupled modes. We separate each mode into its coherent amplitude and a zero-mean fluctuation operator:
\begin{equation}
\begin{aligned}
\hat a &= \alpha+\delta\hat a, \\
\hat m &= \beta+\delta\hat m, \\
\langle\delta\hat a\rangle &= \langle\delta\hat m\rangle=0.
\end{aligned}
\label{eq:appendix_displacement}
\end{equation}
Substituting Eq.~\eqref{eq:appendix_displacement} into Eq.~\eqref{eq:cavity_langevin} and Eq.~\eqref{eq:magnon_langevin}, and then taking expectation values, gives the steady-state first-moment equations
\begin{equation}
\begin{aligned}
0 &= -\lambda_a\alpha-ig\beta+\Omega, \\
0 &= -\lambda_m\beta-ig\alpha.
\end{aligned}
\label{eq:appendix_first_moment_equations}
\end{equation}
Using the complex rates defined below Eq.~\eqref{eq:cavity_amplitude}, the solutions are
\begin{equation}
\begin{aligned}
\alpha &= \frac{\Omega\lambda_m}{\lambda}, \\
\beta &= -\frac{ig\Omega}{\lambda}.
\end{aligned}
\label{eq:appendix_first_moment_solutions}
\end{equation}
The cavity solution in Eq.~\eqref{eq:appendix_first_moment_solutions} reproduces Eq.~\eqref{eq:cavity_amplitude}.

After subtracting the steady-state amplitudes, the fluctuation operators obey
\begin{equation}
\begin{aligned}
\delta\dot{\hat a} &= -\lambda_a\delta\hat a-ig\delta\hat m+\sqrt{2\kappa_a}\,\hat a^{\mathrm{in}}, \\
\delta\dot{\hat m} &= -\lambda_m\delta\hat m-ig\delta\hat a+\sqrt{2\kappa_m}\,\hat m^{\mathrm{in}}.
\end{aligned}
\label{eq:appendix_fluctuation_equations}
\end{equation}
The thermal inputs in Eq.~\eqref{eq:noise_correlations} contain no phase-sensitive pair correlations. The steady-state equations for $\langle\delta\hat a\delta\hat a\rangle$, $\langle\delta\hat m\delta\hat m\rangle$, and $\langle\delta\hat a\delta\hat m\rangle$ therefore form the homogeneous system
\begin{equation}
\begin{aligned}
0 &= -2\lambda_a\langle\delta\hat a\delta\hat a\rangle-2ig\langle\delta\hat a\delta\hat m\rangle, \\
0 &= -2\lambda_m\langle\delta\hat m\delta\hat m\rangle-2ig\langle\delta\hat a\delta\hat m\rangle, \\
0 &= -(\lambda_a+\lambda_m)\langle\delta\hat a\delta\hat m\rangle-ig\left(\langle\delta\hat a\delta\hat a\rangle+\langle\delta\hat m\delta\hat m\rangle\right).
\end{aligned}
\label{eq:anomalous_moment_system}
\end{equation}
For the stable passive system considered here, Eq.~\eqref{eq:anomalous_moment_system} has the unique solution
\begin{equation}
\begin{aligned}
\langle\delta\hat a\delta\hat a\rangle &= \langle\delta\hat m\delta\hat m\rangle=\langle\delta\hat a\delta\hat m\rangle=0, \\
\langle\delta\hat a^{\dagger}\delta\hat a^{\dagger}\rangle &= \langle\delta\hat m^{\dagger}\delta\hat m^{\dagger}\rangle=\langle\delta \hat a^\dagger\delta \hat m^\dagger\rangle=0.
\end{aligned}
\label{eq:phase_sensitive_moments}
\end{equation}
Thus, the passive exchange interaction and thermal inputs do not generate phase-sensitive second-order correlations.

We next define the cavity and magnon fluctuation occupations as $N_a=\langle\delta \hat a^\dagger\delta\hat a\rangle$ and $N_m=\langle\delta \hat m^\dagger\delta\hat m\rangle$, together with the cross correlation $S=\langle\delta \hat a^\dagger\delta\hat m\rangle$. Since the two mode operators commute, $S^*=\langle\delta\hat a\delta \hat m^\dagger\rangle$. Applying Eq.~\eqref{eq:appendix_fluctuation_equations} to these moments gives
\begin{equation}
\begin{aligned}
0 &= -2\kappa_aN_a-igS+igS^*+2\kappa_an_a^{\mathrm{th}}, \\
0 &= -2\kappa_mN_m+igS-igS^*+2\kappa_mn_m^{\mathrm{th}}, \\
0 &=-(\kappa+i\Delta_{am})S + ig(N_m-N_a).
\end{aligned}
\label{eq:normal_moment_system}
\end{equation}
The last line of Eq.~\eqref{eq:normal_moment_system} gives
\begin{equation}
S=\frac{ig(N_m-N_a)}{\kappa+i\Delta_{am}}=\frac{g(\Delta_{am}+i\kappa)(N_m-N_a)}{\kappa^2+\Delta_{am}^2}. \label{eq:cross_correlation_solution}
\end{equation}
Substituting Eq.~\eqref{eq:cross_correlation_solution} into the first two lines of Eq.~\eqref{eq:normal_moment_system} yields
\begin{equation}
\begin{aligned}
\kappa_a\left(N_a-n_a^{\mathrm{th}}\right) &= R\left(N_m-N_a\right), \\
\kappa_m\left(N_m-n_m^{\mathrm{th}}\right) &= -R\left(N_m-N_a\right),
\end{aligned}
\label{eq:occupation_balance}
\end{equation}
where the thermal-exchange rate entering these balance equations is $R=g^2\kappa/(\kappa^2+\Delta_{am}^2)$. The opposite signs in Eq.~\eqref{eq:occupation_balance} express the transfer of thermal fluctuations between the two modes while each mode remains coupled to its own reservoir.

Solving Eq.~\eqref{eq:occupation_balance} for the occupation difference gives
\begin{equation}
N_m-N_a=\Delta n^{\mathrm{th}}\frac{\kappa^2+\Delta_{am}^2}{(1+C)\kappa^2+\Delta_{am}^2}. \label{eq:coupled_occupation_difference}
\end{equation}
Substituting Eq.~\eqref{eq:coupled_occupation_difference} into the first line of Eq.~\eqref{eq:occupation_balance} gives $N_a=n_a^{\mathrm{th}}+\eta_m\Delta n^{\mathrm{th}}$, where the coefficient multiplying the occupation imbalance is $g^2\kappa/\{\kappa_a[(1+C)\kappa^2+\Delta_{am}^2]\}=C\kappa_m\kappa/[(1+C)\kappa^2+\Delta_{am}^2]=\eta_m$. This reproduces the magnon-to-cavity thermal transfer coefficient defined in Eq.~\eqref{eq:thermal_transfer}. The cavity fluctuation occupation can therefore be written as
\begin{equation}
N_a=(1-\eta_m)n_a^{\mathrm{th}}+\eta_mn_m^{\mathrm{th}}. \label{eq:appendix_cavity_occupation}
\end{equation}

Finally, the fluctuation of the measured amplitude quadrature is $\delta \hat X_a=\hat X_a-\mu=(\delta\hat a+\delta \hat a^\dagger)/\sqrt{2}$. Using Eq.~\eqref{eq:phase_sensitive_moments}, Eq.~\eqref{eq:appendix_cavity_occupation}, and the commutator $\langle\delta\hat a\delta \hat a^\dagger\rangle=N_a+1$, its variance becomes
\begin{equation}
\begin{aligned}
\sigma^2 &= \langle\delta \hat X_a^2\rangle, \\
&= \frac{1}{2}\left(\langle\delta\hat a\delta\hat a\rangle+\langle\delta\hat a^{\dagger}\delta\hat a^{\dagger}\rangle+\langle\delta \hat a^\dagger\delta\hat a\rangle+\langle\delta\hat a\delta \hat a^\dagger\rangle\right), \\
&= N_a+\frac{1}{2}, \\
&= (1-\eta_m)n_a^{\mathrm{th}}+\eta_mn_m^{\mathrm{th}}+\frac{1}{2}.
\end{aligned}
\label{eq:appendix_cavity_variance}
\end{equation}
Eq.~\eqref{eq:appendix_cavity_variance} reproduces the variance stated in Eq.~\eqref{eq:cavity_variance}. The vacuum term $1/2$ is field independent, while the thermal part depends on the field through both the transfer coefficient $\eta_m$ and the magnon occupation $n_m^{\mathrm{th}}$.

\section{Classical Fisher Information and Field Derivatives}
\label{app:cfi}

We derive the classical Fisher information of the Gaussian likelihood in Eq.~\eqref{eq:gaussian_likelihood} and the field derivatives used throughout the main text. For compactness, we define the centered measurement outcome as $z=x-\mu(B)$. Differentiating the logarithm of the likelihood gives
\begin{equation}
\partial_B\ln p(x|B)=\frac{z}{\sigma^2}\partial_B\mu+\frac{z^2-\sigma^2}{2\sigma^4}\partial_B\sigma^2. \label{eq:appendix_gaussian_score}
\end{equation}
The required centered Gaussian moments are
\begin{equation}
\begin{aligned}
\langle z\rangle_p &= \langle z^3\rangle_p=0, \\
\langle z^2\rangle_p &= \sigma^2, \\
\langle z^4\rangle_p &= 3\sigma^4,
\end{aligned}
\label{eq:appendix_gaussian_moments}
\end{equation}
where $\langle\cdot\rangle_p$ denotes an average over $p(x|B)$. Substituting Eq.~\eqref{eq:appendix_gaussian_score} into the Fisher-information definition in Eq.~\eqref{eq:cfi_definition} and using Eq.~\eqref{eq:appendix_gaussian_moments} gives
\begin{equation}
\begin{aligned}
\mathcal{F}(B) &= \frac{(\partial_B\mu)^2}{\sigma^4}\langle z^2\rangle_p+\frac{(\partial_B\mu)(\partial_B\sigma^2)}{\sigma^6}\langle z(z^2-\sigma^2)\rangle_p \\
&+\frac{(\partial_B\sigma^2)^2}{4\sigma^8}\langle(z^2-\sigma^2)^2\rangle_p, \\
&= \frac{(\partial_B\mu)^2}{\sigma^2}+\frac{(\partial_B\sigma^2)^2}{2\sigma^4}.
\end{aligned}
\label{eq:appendix_cfi_decomposition}
\end{equation}
The cross term in Eq.~\eqref{eq:appendix_cfi_decomposition} vanishes because it contains only odd centered moments, while $\langle(z^2-\sigma^2)^2\rangle_p=2\sigma^4$. Eq.~\eqref{eq:appendix_cfi_decomposition} reproduces the exact decomposition in Eq.~\eqref{eq:cfi_decomposition}.

For the local field derivative, we treat $g$, $\kappa_a$, $\kappa_m$, $\omega_a$, and $\Omega$ as independent of $B$. The field enters the steady state through $\Delta_m(B)$ and $n_m^{\mathrm{th}}(B)$, with $\partial_B\Delta_m=-\gamma$. We first evaluate the derivative of the coherent amplitude in Eq.~\eqref{eq:cavity_amplitude}. Since $\partial_{\Delta_m}\lambda_m=-i$ and $\partial_{\Delta_m}\lambda=-i\lambda_a$, direct differentiation gives
\begin{equation}
\frac{\partial\alpha}{\partial\Delta_m}=\frac{\Omega[-i\lambda+i\lambda_a\lambda_m]}{\lambda^2}=-\frac{i\Omega g^2}{\lambda^2}. \label{eq:appendiX_alpha_derivative}
\end{equation}
Using $\lambda=A-iD$ from Eq.~\eqref{eq:A_D_definitions}, the real part of Eq.~\eqref{eq:appendiX_alpha_derivative} gives
\begin{equation}
\begin{aligned}
\frac{\partial\mu}{\partial\Delta_m} &= \frac{2\sqrt{2}\,\Omega g^2AD}{(A^2+D^2)^2}, \\
\frac{\partial\mu}{\partial B} &= -\gamma\frac{\partial\mu}{\partial\Delta_m}.
\end{aligned}
\label{eq:appendix_mean_derivatives}
\end{equation}
Eq.~\eqref{eq:appendix_mean_derivatives} reproduces the corrected field derivative in Eq.~\eqref{eq:mean_field_derivative}. In particular, the numerator of $\partial_{\Delta_m}\mu$ is proportional to $AD$; the factor $A^2D^2$ appears only after this derivative is squared in $\mathcal{F}_{\rm coh}$.

The dimensionless form follows from the normalized detunings defined in Eq.~\eqref{eq:normalized_detunings} and the cooperativity $C=g^2/(\kappa_a\kappa_m)$. The real and imaginary parts of $\lambda$ become
\begin{equation}
\begin{aligned}
A &= \kappa_a\kappa_m(1+C-\delta_a\delta_m), \\
D &= \kappa_a\kappa_m(\delta_a+\delta_m), \\
\frac{A^2+D^2}{\kappa_a^2\kappa_m^2} &= (1+C-\delta_a\delta_m)^2+(\delta_a+\delta_m)^2.
\end{aligned}
\label{eq:appendix_dimensionless_AD}
\end{equation}
Substituting Eq.~\eqref{eq:appendix_dimensionless_AD} into the first term of Eq.~\eqref{eq:appendix_cfi_decomposition} gives the coherent factorization in Eq.~\eqref{eq:coherent_cfi_factorization} and the morphology in Eq.~\eqref{eq:coherent_morphology}.

We next derive the field dependence of the cavity variance. Eq.~\eqref{eq:appendix_cavity_variance} can be written as $\sigma^2=n_a^{\mathrm{th}}+\eta_m\Delta n^{\mathrm{th}}+1/2$. At fixed temperature and cavity frequency, $n_a^{\mathrm{th}}$ is independent of $B$. Differentiating the Bose occupation and using the thermal susceptibility defined in Eq.~\eqref{eq:thermal_susceptibility} gives
\begin{equation}
\begin{aligned}
\frac{\partial n_m^{\mathrm{th}}}{\partial\omega_m} &= -\xi_m, \\
\frac{\partial n_m^{\mathrm{th}}}{\partial\Delta_m} &= \xi_m, \\
\frac{\partial n_m^{\mathrm{th}}}{\partial B} &= -\gamma\xi_m.
\end{aligned}
\label{eq:appendix_thermal_occupation_derivatives}
\end{equation}
Because $\partial_{\Delta_m}\Delta_{am}=-1$, differentiating the transfer coefficient in Eq.~\eqref{eq:thermal_transfer} gives
\begin{equation}
\frac{\partial\eta_m}{\partial\Delta_m}=\frac{2C\kappa_m\kappa\Delta_{am}}{[(1+C)\kappa^2+\Delta_{am}^2]^2}=\frac{2\kappa_a\Delta_{am}}{g^2\kappa}\eta_m^2. \label{eq:appendix_transfer_derivative}
\end{equation}
Applying Eq.~\eqref{eq:appendix_thermal_occupation_derivatives} and Eq.~\eqref{eq:appendix_transfer_derivative} to the variance gives
\begin{equation}
\partial_B\sigma^2=-\gamma\left[\Delta n^{\mathrm{th}}\frac{\partial\eta_m}{\partial\Delta_m}+\eta_m\xi_m\right]. \label{eq:appendix_variance_derivative}
\end{equation}
Eq.~\eqref{eq:appendix_variance_derivative} reproduces Eq.~\eqref{eq:variance_field_derivative}. Its first term results from the field dependence of the thermal-transfer coefficient, while its second term results from the field dependence of the magnon thermal occupation.

To obtain the dimensionless form, we use $x_{\rm eff}$ from Eq.~\eqref{eq:effective_detuning} and define $Q=1+C+x_{\rm eff}^2$. The transfer coefficient becomes $\eta_m=C\kappa_m/(\kappa Q)$. Eq.~\eqref{eq:appendix_variance_derivative} then reduces to
\begin{equation}
\partial_B\sigma^2=-\gamma\frac{C\kappa_m}{\kappa^2}\frac{2x_{\rm eff}\Delta n^{\mathrm{th}}+\xi_m\kappa Q}{Q^2}. \label{eq:appendix_dimensionless_variance_derivative}
\end{equation}
Squaring Eq.~\eqref{eq:appendix_dimensionless_variance_derivative} and substituting it into the second term of Eq.~\eqref{eq:appendix_cfi_decomposition} gives the exact thermal factorization in Eq.~\eqref{eq:incoherent_cfi_factorization} and the response function in Eq.~\eqref{eq:exact_incoherent_morphology}.

Finally, within the local thermal expansion in Eq.~\eqref{eq:local_thermal_expansion}, the numerator of Eq.~\eqref{eq:appendix_dimensionless_variance_derivative} becomes $\xi_a\kappa(1+C-x_{\rm eff}^2)$. The variance derivative therefore takes the reduced form
\begin{equation}
\partial_B\sigma^2\simeq-\gamma\frac{\xi_aC\kappa_m}{\kappa}\frac{1+C-x_{\rm eff}^2}{(1+C+x_{\rm eff}^2)^2}. \label{eq:appendix_reduced_variance_derivative}
\end{equation}
Substitution of Eq.~\eqref{eq:appendix_reduced_variance_derivative} into Eq.~\eqref{eq:appendix_cfi_decomposition} gives the reduced thermal Fisher information in Eq.~\eqref{eq:reduced_incoherent_cfi} and the morphology in Eq.~\eqref{eq:reduced_incoherent_morphology}.

\section{Global Optimization of the Coherent Morphology}
\label{app:coherent_optimization}

We determine the global maximum of the coherent morphology in Eq.~\eqref{eq:coherent_morphology} over $C>0$ and real normalized detunings. We first show that every nonzero detuning maximum lies on the symmetric line $\delta_a=\delta_m$ and then optimize the resulting reduced function over $C$ and the common detuning.

For compactness, we write $x=\delta_a$ and $y=\delta_m$ and introduce
\begin{equation}
\begin{aligned}
s &= x+y, \\
q &= 1+C-xy, \\
r^2 &= q^2+s^2.
\end{aligned}
\label{eq:appendix_coherent_variables}
\end{equation}
The coherent morphology then becomes
\begin{equation}
f_{\rm coh}=\frac{4C^2s^2q^2}{r^8}. \label{eq:appendix_coherent_compact}
\end{equation}
The cases $s=0$ and $q=0$ give $f_{\rm coh}=0$ and therefore cannot be global maxima because positive values exist elsewhere. For $sq\neq0$, the stationary points of $f_{\rm coh}$ are also the stationary points of $\Phi=\ln f_{\rm coh}$. Differentiating Eq.~\eqref{eq:appendix_coherent_compact} with respect to $x$ and $y$ gives
\begin{equation}
\begin{aligned}
\frac{\partial\Phi}{\partial x} &= \frac{2}{s}-\frac{2y}{q}+\frac{8qy}{r^2}-\frac{8s}{r^2}, \\
\frac{\partial\Phi}{\partial y} &= \frac{2}{s}-\frac{2x}{q}+\frac{8qx}{r^2}-\frac{8s}{r^2}.
\end{aligned}
\label{eq:appendix_coherent_log_derivatives}
\end{equation}
At a stationary point, both lines of Eq.~\eqref{eq:appendix_coherent_log_derivatives} vanish. Subtracting the second line from the first gives
\begin{equation}
\frac{\partial\Phi}{\partial x}-\frac{\partial\Phi}{\partial y}=\frac{2(x-y)(s^2-3q^2)}{qr^2}=0. \label{eq:appendix_coherent_derivative_difference}
\end{equation}
Eq.~\eqref{eq:appendix_coherent_derivative_difference} allows either $x=y$ or $s^2=3q^2$. If the second condition is inserted into either line of Eq.~\eqref{eq:appendix_coherent_log_derivatives}, the remaining stationary condition requires $s^2=q^2$. The two relations can hold simultaneously only when $s=q=0$, which belongs to the zero-information set excluded above. Every nonzero stationary maximum must therefore satisfy $x=y$, proving the symmetry relation in Eq.~\eqref{eq:coherent_symmetry_line}.

We now set $x=y=\delta_{\rm sym}$ and use the reduced morphology in Eq.~\eqref{eq:coherent_reduced_morphology}. To optimize it jointly over $C$ and $\delta_{\rm sym}$, we introduce the polar variables
\begin{equation}
\begin{aligned}
1+C-\delta_{\rm sym}^2 &= \rho\cos\theta, \\
2\delta_{\rm sym} &= \rho\sin\theta, \\
C &= \rho\cos\theta-1+\frac{\rho^2\sin^2\theta}{4},
\end{aligned}
\label{eq:appendix_coherent_polar_variables}
\end{equation}
where $\rho>0$. Substitution of Eq.~\eqref{eq:appendix_coherent_polar_variables} into Eq.~\eqref{eq:coherent_reduced_morphology} gives
\beqa
    f_{\rm coh}&=&\sin^2(2\theta)\left(\frac{C}{\rho^2}\right)^2 \notag \\
&=&\sin^2(2\theta)\left(\frac{\cos\theta}{\rho}-\frac{1}{\rho^2}+\frac{\sin^2\theta}{4}\right)^2. \label{eq:appendix_coherent_polar_morphology}
\eeqa

Defining $u=1/\rho$, the factor containing the radial variable can be written as
\begin{equation}
\frac{C}{\rho^2}=u\cos\theta-u^2+\frac{\sin^2\theta}{4}=\frac{1}{4}-\left(u-\frac{\cos\theta}{2}\right)^2\leq\frac{1}{4}. \label{eq:appendix_coherent_radial_bound}
\end{equation}
Using Eq.~\eqref{eq:appendix_coherent_radial_bound} in Eq.~\eqref{eq:appendix_coherent_polar_morphology} gives the global bound
\begin{equation}
f_{\rm coh}\leq\frac{\sin^2(2\theta)}{16}\leq\frac{1}{16}. \label{eq:appendix_coherent_global_bound}
\end{equation}
Equality in Eq.~\eqref{eq:appendix_coherent_global_bound} requires
\begin{equation}
\begin{aligned}
u &= \frac{\cos\theta}{2}, \\
\sin^2(2\theta) &= 1, \\
\cos\theta &> 0.
\end{aligned}
\label{eq:appendix_coherent_equality_conditions}
\end{equation}
The last condition follows from $u=1/\rho>0$. Solving Eq.~\eqref{eq:appendix_coherent_equality_conditions} and transforming back with Eq.~\eqref{eq:appendix_coherent_polar_variables} gives
\begin{equation}
\begin{aligned}
\theta_{\rm opt} &= \pm\frac{\pi}{4}, \\
\rho_{\rm opt} &= 2\sqrt{2}, \\
\delta_{\rm sym, opt} &= \pm1, \\
C_{\rm opt} &= 2, \\
f_{\rm coh}^{\rm max} &= \frac{1}{16}.
\end{aligned}
\label{eq:appendix_coherent_global_optimum}
\end{equation}
The signs of $\theta_{\rm opt}$ and $\delta_{\rm sym, opt}$ are correlated. Eq.~\eqref{eq:appendix_coherent_global_optimum} proves the optimum stated in Eq.~\eqref{eq:coherent_global_optimum}. With fixed damping rates, substituting $C_{\rm opt}=2$ into the definition of cooperativity gives the coupling in Eq.~\eqref{eq:optimal_coupling}.

This optimization applies to the dimensionless coherent morphology of the chosen cavity amplitude quadrature under the constant-variance approximation stated in the main text. The proof does not establish the same optimum for a different homodyne angle or for a parameter change in which the damping rates vary together with the coupling.

\section{Stationary Structure of the Thermal-Fluctuation Morphology}
\label{app:incoherent_optimization}

We classify the stationary points of the reduced thermal morphology in Eq.~\eqref{eq:reduced_incoherent_morphology}. For fixed $C>0$, we write $x=x_{\rm eff}$ and define $a=1+C$. The morphology then becomes
\begin{equation}
\widetilde f_{\rm th}(C,x)=\frac{C^2(a-x^2)^2}{(a+x^2)^4}. \label{eq:appendix_incoherent_compact}
\end{equation}
The function in Eq.~\eqref{eq:appendix_incoherent_compact} is even in $x$ and approaches zero as $|x|\rightarrow\infty$. Its derivative is
\begin{equation}
\frac{\partial\widetilde f_{\rm th}}{\partial x}=-\frac{4C^2x(a-x^2)(3a-x^2)}{(a+x^2)^5}. \label{eq:appendix_incoherent_derivative}
\end{equation}
Eq.~\eqref{eq:appendix_incoherent_derivative} is equivalent to Eq.~\eqref{eq:incoherent_detuning_derivative}. Its stationary points satisfy
\begin{equation}
\begin{aligned}
x &= 0, \\
x^2 &= a, \\
x^2 &= 3a.
\end{aligned}
\label{eq:appendix_incoherent_stationary_points}
\end{equation}

For $x>0$, the derivative in Eq.~\eqref{eq:appendix_incoherent_derivative} is negative over $0<x<\sqrt{a}$, positive over $\sqrt{a}<x<\sqrt{3a}$, and negative again for $x>\sqrt{3a}$. The symmetry of the function gives the corresponding behavior for $x<0$. The stationary values are
\begin{equation}
\begin{aligned}
\widetilde f_{\rm th}(C,0) &= \frac{C^2}{a^2}, \\
\widetilde f_{\rm th}(C,\pm\sqrt{a}) &= 0, \\
\widetilde f_{\rm th}(C,\pm\sqrt{3a}) &= \frac{C^2}{64a^2}.
\end{aligned}
\label{eq:appendix_incoherent_stationary_values}
\end{equation}
Eq.~\eqref{eq:appendix_incoherent_stationary_values} shows that $x=0$ is a local maximum, $x^2=a$ gives two zero-valued minima, and $x^2=3a$ gives two side maxima. Because the function approaches zero at large $|x|$ and the side maxima are smaller than the central one, $x=0$ is the global maximum. The relative height of the side maxima is
\begin{equation}
\frac{\widetilde f_{\rm th}(C,\pm\sqrt{3a})}{\widetilde f_{\rm th}(C,0)}=\frac{1}{64}. \label{eq:appendix_incoherent_side_ratio}
\end{equation}
Eq.~\eqref{eq:appendix_incoherent_side_ratio} explains why the central ridge dominates Fig.~\ref{fig:thermal_dispersion}, while also showing that the morphology does not decrease strictly at every point away from $x=0$.

Restoring $a=1+C$, the global maximum over relative detuning is the central value given in Eq.~\eqref{eq:optimized_incoherent_morphology}. Its derivative with respect to cooperativity is
\begin{equation}
\frac{d\widetilde f_{\rm th}^{\rm opt}}{dC}=\frac{d}{dC}\left[\frac{C^2}{(1+C)^2}\right]=\frac{2C}{(1+C)^3}>0. \label{eq:appendix_incoherent_C_derivative}
\end{equation}
Eq.~\eqref{eq:appendix_incoherent_C_derivative} shows that the optimized reduced morphology increases with $C$ for every $C>0$. It approaches zero as $C\rightarrow0$ and the finite value $1$ as $C\rightarrow\infty$.

The stationary-point classification applies to the reduced morphology obtained from the local thermal expansion in Eq.~\eqref{eq:local_thermal_expansion}. The increase with $C$ concerns only $\widetilde f_{\rm th}^{\rm opt}$; the dimensional prefactor in Eq.~\eqref{eq:reduced_incoherent_cfi} and the range of validity of the adopted dissipative model must also be considered when assessing the complete Fisher information.

\section{Sensitivity analysis}
\label{app:sensitivity}
{  For a single statistically independent quadrature outcome, let $\mathcal{F}_{\rm total}$ denote the Fisher information about the magnetic field. Because Fisher information is additive for independent measurements, $N$ independent outcomes contain the total information $N\mathcal{F}_{\rm total}$. The Cram\'er--Rao bound therefore gives
\begin{equation}
\delta B\equiv\sqrt{\operatorname{Var}(\widehat B)}\geq\delta B_{\rm CR}\equiv\frac{1}{\sqrt{N\mathcal{F}_{\rm total}}}.
\label{eq:CR_independent}
\end{equation}

For a continuous quadrature record, temporally adjacent outcomes are correlated because the cavity--magnon system has a finite relaxation time. Introducing an effective correlation time $\tau_{\rm corr}$, a record of duration $\tau$ contains approximately $N_{\rm eff}\simeq\tau/\tau_{\rm corr}$ statistically independent outcomes. Substitution of $N_{\rm eff}$ into Eq.~\eqref{eq:CR_independent} gives
\begin{equation} 
\delta B_{\rm CR}(\tau)=\sqrt{\frac{\tau_{\rm corr}}{\tau\,\mathcal{F}_{\rm total}}},\quad \mathcal{S}_{\rm total}\equiv\delta B_{\rm CR}(\tau)\sqrt{\tau}=\sqrt{\frac{\tau_{\rm corr}}{\mathcal{F}_{\rm total}}}.
\label{eq:app_sensitivity_rate}
\end{equation}
The estimation uncertainty therefore satisfies $\delta B(\tau)\geq\delta B_{\rm CR}(\tau)$ and $\delta B(\tau)\sqrt{\tau}\geq\mathcal{S}_{\rm total}$. When the field uncertainty scales as $\tau^{-1/2}$, the time-normalized quantity $\delta B(\tau)\sqrt{\tau}$ characterizes the magnetic-field sensitivity. Its units are $\mathrm{T}\sqrt{\mathrm{s}}=\mathrm{T}/\sqrt{\mathrm{Hz}}$. The quantity $\mathcal{S}_{\rm total}$ is the Cram\'er--Rao lower bound on this sensitivity and is reached when the estimator saturates the uncertainty bound. A smaller $\mathcal{S}_{\rm total}$ corresponds to a lower uncertainty at a given integration time.}

\bibliography{Reference}

@article{aspelmeyer2014cavity,
  title = {Cavity optomechanics},
  author = {Aspelmeyer, Markus and Kippenberg, Tobias J. and Marquardt, Florian},
  year = {2014},
  month = dec,
  journal = {Rev. Mod. Phys.},
  volume = {86},
  number = {4},
  pages = {1391--1452},
  doi = {10.1103/RevModPhys.86.1391}
}

@article{bai2015spin,
  title = {Spin Pumping in Electrodynamically Coupled Magnon-Photon Systems},
  author = {Bai, Lihui and Harder, M. and Chen, Y. P. and Fan, X. and Xiao, J. Q. and Hu, C.-M.},
  year = {2015},
  month = jun,
  journal = {Phys. Rev. Lett.},
  volume = {114},
  number = {22},
  pages = {227201},
  doi = {10.1103/PhysRevLett.114.227201}
}

@article{blais2021circuit,
  title = {Circuit quantum electrodynamics},
  author = {Blais, Alexandre and Grimsmo, Arne L. and Girvin, S. M. and Wallraff, Andreas},
  year = {2021},
  month = may,
  journal = {Rev. Mod. Phys.},
  volume = {93},
  number = {2},
  pages = {025005},
  doi = {10.1103/RevModPhys.93.025005}
}

@article{braggio2017optical,
  title = {Optical Manipulation of a Magnon-Photon Hybrid System},
  author = {Braggio, C. and Carugno, G. and Guarise, M. and Ortolan, A. and Ruoso, G.},
  year = {2017},
  month = mar,
  journal = {Phys. Rev. Lett.},
  volume = {118},
  number = {10},
  pages = {107205},
  doi = {10.1103/PhysRevLett.118.107205}
}

@article{cao2019exceptional,
  title = {Exceptional magnetic sensitivity of P T -symmetric cavity magnon polaritons},
  author = {Cao, Yunshan and Yan, Peng},
  year = {2019},
  month = jun,
  journal = {Phys. Rev. B},
  volume = {99},
  number = {21},
  pages = {214415},
  doi = {10.1103/PhysRevB.99.214415}
}

@article{clerk2010introduction,
  title = {Introduction to quantum noise, measurement, and amplification},
  author = {Clerk, A. A. and Devoret, M. H. and Girvin, S. M. and Marquardt, Florian and Schoelkopf, R. J.},
  year = {2010},
  month = apr,
  journal = {Rev. Mod. Phys.},
  volume = {82},
  number = {2},
  pages = {1155--1208},
  doi = {10.1103/RevModPhys.82.1155}
}

@article{crescini2020cavity,
  title = {Cavity magnon polariton based precision magnetometry},
  author = {Crescini, N. and Braggio, C. and Carugno, G. and Ortolan, A. and Ruoso, G.},
  year = {2020},
  month = oct,
  journal = {Appl. Phys. Lett.},
  volume = {117},
  number = {14},
  pages = {144001},
  doi = {10.1063/5.0024369}
}

@article{crescini2021phasemodulated,
  title = {Phase-Modulated Cavity Magnon Polaritons as a Precise Magnetic Field Probe},
  author = {Crescini, Nicol{\`o} and Carugno, Giovanni and Ruoso, Giuseppe},
  year = {2021},
  month = sep,
  journal = {Phys. Rev. Applied},
  volume = {16},
  number = {3},
  pages = {034036},
  doi = {10.1103/PhysRevApplied.16.034036}
}

@article{ebrahimi2021singlequadrature,
  title = {Single-quadrature quantum magnetometry in cavity electromagnonics},
  author = {Ebrahimi, M. S. and Motazedifard, Ali and Harouni, M. Bagheri},
  year = {2021},
  month = jun,
  journal = {Phys. Rev. A},
  volume = {103},
  number = {6},
  pages = {062605},
  doi = {10.1103/PhysRevA.103.062605}
}

@article{fan2022optical,
  title = {Optical sensing of magnons via the magnetoelastic displacement},
  author = {Fan, Zhi-Yuan and Shen, Rui-Chang and Wang, Yi-Pu and Li, Jie and You, J. Q.},
  year = {2022},
  month = mar,
  journal = {Phys. Rev. A},
  volume = {105},
  number = {3},
  pages = {033507},
  doi = {10.1103/PhysRevA.105.033507}
}

@article{flower2019experimental,
  title = {Experimental implementations of cavity-magnon systems: from ultra strong coupling to applications in precision measurement},
  author = {Flower, Graeme and Goryachev, Maxim and Bourhill, Jeremy and Tobar, Michael E},
  year = {2019},
  month = sep,
  journal = {New J. Phys.},
  volume = {21},
  number = {9},
  pages = {095004},
  doi = {10.1088/1367-2630/ab3e1c}
}

@article{gardiner1985input,
  title = {Input and output in damped quantum systems: Quantum stochastic differential equations and the master equation},
  author = {Gardiner, C. W. and Collett, M. J.},
  year = {1985},
  month = jun,
  journal = {Phys. Rev. A},
  volume = {31},
  number = {6},
  pages = {3761--3774},
  doi = {10.1103/PhysRevA.31.3761}
}

@article{goryachev2014highcooperativity,
  title = {High-Cooperativity Cavity QED with Magnons at Microwave Frequencies},
  author = {Goryachev, Maxim and Farr, Warrick G. and Creedon, Daniel L. and Fan, Yaohui and Kostylev, Mikhail and Tobar, Michael E.},
  year = {2014},
  month = nov,
  journal = {Phys. Rev. Appl.},
  volume = {2},
  number = {5},
  pages = {054002},
  doi = {10.1103/PhysRevApplied.2.054002}
}

@article{gui2026gaindriven,
  title = {Gain-driven polaritons for high-sensitivity wireless magnetic sensing},
  author = {Gui, Y. S. and Hu, C.-M.},
  year = {2026},
  month = jun,
  journal = {APL Quantum},
  volume = {3},
  number = {2},
  pages = {026110},
  doi = {10.1063/5.0331021}
}

@article{haigh2016tripleresonant,
  title = {Triple-Resonant Brillouin Light Scattering in Magneto-Optical Cavities},
  author = {Haigh, J. A. and Nunnenkamp, A. and Ramsay, A. J. and Ferguson, A. J.},
  year = {2016},
  month = sep,
  journal = {Phys. Rev. Lett.},
  volume = {117},
  number = {13},
  pages = {133602},
  doi = {10.1103/PhysRevLett.117.133602}
}

@article{hisatomi2016bidirectional,
  title = {Bidirectional conversion between microwave and light via ferromagnetic magnons},
  author = {Hisatomi, R. and Osada, A. and Tabuchi, Y. and Ishikawa, T. and Noguchi, A. and Yamazaki, R. and Usami, K. and Nakamura, Y.},
  year = {2016},
  month = may,
  journal = {Phys. Rev. B},
  volume = {93},
  number = {17},
  pages = {174427},
  doi = {10.1103/PhysRevB.93.174427}
}

@article{huebl2013high,
  title = {High Cooperativity in Coupled Microwave Resonator Ferrimagnetic Insulator Hybrids},
  author = {Huebl, Hans and Zollitsch, Christoph W. and Lotze, Johannes and Hocke, Fredrik and Greifenstein, Moritz and Marx, Achim and Gross, Rudolf and Goennenwein, Sebastian T. B.},
  year = {2013},
  month = sep,
  journal = {Phys. Rev. Lett.},
  volume = {111},
  number = {12},
  pages = {127003},
  doi = {10.1103/PhysRevLett.111.127003}
}

@article{klingler2016combined,
  title = {Combined Brillouin light scattering and microwave absorption study of magnon-photon coupling in a split-ring resonator/YIG film system},
  author = {Klingler, S. and {Maier-Flaig}, H. and Gross, R. and Hu, C.-M. and Huebl, H. and Goennenwein, S. T. B. and Weiler, M.},
  year = {2016},
  month = aug,
  journal = {Appl. Phys. Lett.},
  volume = {109},
  number = {7},
  pages = {072402},
  doi = {10.1063/1.4961052}
}

@article{lachance-quirion2019hybrid,
  title = {Hybrid quantum systems based on magnonics},
  author = {{Lachance-Quirion}, Dany and Tabuchi, Yutaka and Gloppe, Arnaud and Usami, Koji and Nakamura, Yasunobu},
  year = {2019},
  month = jun,
  journal = {Appl. Phys. Express},
  volume = {12},
  number = {7},
  pages = {070101},
  doi = {10.7567/1882-0786/ab248d}
}

@article{langbein2018no,
  title = {No exceptional precision of exceptional-point sensors},
  author = {Langbein, W.},
  year = {2018},
  month = aug,
  journal = {Phys. Rev. A},
  volume = {98},
  number = {2},
  pages = {023805},
  doi = {10.1103/PhysRevA.98.023805}
}

@article{li2018magnonphotonphonon,
  title = {Magnon-Photon-Phonon Entanglement in Cavity Magnomechanics},
  author = {Li, Jie and Zhu, Shi-Yao and Agarwal, G. S.},
  year = {2018},
  month = nov,
  journal = {Phys. Rev. Lett.},
  volume = {121},
  number = {20},
  pages = {203601},
  doi = {10.1103/PhysRevLett.121.203601}
}

@article{li2019strong,
  title = {Strong Coupling between Magnons and Microwave Photons in On-Chip Ferromagnet-Superconductor Thin-Film Devices},
  author = {Li, Yi and Polakovic, Tomas and Wang, Yong-Lei and Xu, Jing and Lendinez, Sergi and Zhang, Zhizhi and Ding, Junjia and Khaire, Trupti and Saglam, Hilal and Divan, Ralu and Pearson, John and Kwok, Wai-Kwong and Xiao, Zhili and Novosad, Valentine and Hoffmann, Axel and Zhang, Wei},
  year = {2019},
  month = sep,
  journal = {Phys. Rev. Lett.},
  volume = {123},
  number = {10},
  pages = {107701},
  doi = {10.1103/PhysRevLett.123.107701}
}

@article{li2020hybrid,
  title = {Hybrid magnonics: Physics, circuits, and applications for coherent information processing},
  author = {Li, Yi and Zhang, Wei and Tyberkevych, Vasyl and Kwok, Wai-Kwong and Hoffmann, Axel and Novosad, Valentine},
  year = {2020},
  month = oct,
  journal = {J. Appl. Phys.},
  volume = {128},
  number = {13},
  pages = {130902},
  doi = {10.1063/5.0020277}
}

@article{liu2026quantum,
  title = {Quantum magnetometry with orientation beyond steady-state limits in cavity-magnon systems},
  author = {Liu, Zheng and Xu, Ding-hui and Yang, Yi-jia and Liu, Yu-qiang and Yu, Chang-shui},
  year = {2026},
  month = jun,
  journal = {Phys. Rev. Lett.},
  volume = {137},
  number = {10},
  pages = {103602},
  doi = {10.1103/z982-mv6k}
}

@article{nielsen2025variational,
  title = {Variational quantum algorithm for enhanced continuous variable optical phase sensing},
  author = {Nielsen, Jens A. H. and Kicinski, Mateusz J. and Arge, Tummas N. and Vijayadharan, Kannan and Foldager, Jonathan and Borregaard, Johannes and Meyer, Johannes Jakob and {Neergaard-Nielsen}, Jonas S. and Gehring, Tobias and Andersen, Ulrik L.},
  year = {2025},
  month = may,
  journal = {npj Quantum Inf},
  volume = {11},
  number = {1},
  pages = {70},
  doi = {10.1038/s41534-024-00947-1}
}

@article{osada2016cavity,
  title = {Cavity Optomagnonics with Spin-Orbit Coupled Photons},
  author = {Osada, A. and Hisatomi, R. and Noguchi, A. and Tabuchi, Y. and Yamazaki, R. and Usami, K. and Sadgrove, M. and Yalla, R. and Nomura, M. and Nakamura, Y.},
  year = {2016},
  month = jun,
  journal = {Phys. Rev. Lett.},
  volume = {116},
  number = {22},
  pages = {223601},
  doi = {10.1103/PhysRevLett.116.223601}
}

@article{osada2018brillouin,
  title = {Brillouin Light Scattering by Magnetic Quasivortices in Cavity Optomagnonics},
  author = {Osada, A. and Gloppe, A. and Hisatomi, R. and Noguchi, A. and Yamazaki, R. and Nomura, M. and Nakamura, Y. and Usami, K.},
  year = {2018},
  month = mar,
  journal = {Phys. Rev. Lett.},
  volume = {120},
  number = {13},
  pages = {133602},
  doi = {10.1103/PhysRevLett.120.133602}
}

@article{rumyantsev2019discrete,
  title = {The discrete noise of magnons},
  author = {Rumyantsev, S. and Balinskiy, M. and Kargar, F. and Khitun, A. and Balandin, A. A.},
  year = {2019},
  month = mar,
  journal = {Appl. Phys. Lett.},
  volume = {114},
  number = {9},
  pages = {090601},
  doi = {10.1063/1.5088651}
}

@article{shen2022coherent,
  title = {Coherent Coupling between Phonons, Magnons, and Photons},
  author = {Shen, Zhen and Xu, Guan-Ting and Zhang, Mai and Zhang, Yan-Lei and Wang, Yu and Chai, Cheng-Zhe and Zou, Chang-Ling and Guo, Guang-Can and Dong, Chun-Hua},
  year = {2022},
  month = dec,
  journal = {Phys. Rev. Lett.},
  volume = {129},
  number = {24},
  pages = {243601},
  doi = {10.1103/PhysRevLett.129.243601}
}

@article{tabuchi2014hybridizing,
  title = {Hybridizing Ferromagnetic Magnons and Microwave Photons in the Quantum Limit},
  author = {Tabuchi, Yutaka and Ishino, Seiichiro and Ishikawa, Toyofumi and Yamazaki, Rekishu and Usami, Koji and Nakamura, Yasunobu},
  year = {2014},
  month = aug,
  journal = {Phys. Rev. Lett.},
  volume = {113},
  number = {8},
  pages = {083603},
  doi = {10.1103/PhysRevLett.113.083603}
}

@article{tabuchi2015coherent,
  title = {Coherent coupling between a ferromagnetic magnon and a superconducting qubit},
  author = {Tabuchi, Yutaka and Ishino, Seiichiro and Noguchi, Atsushi and Ishikawa, Toyofumi and Yamazaki, Rekishu and Usami, Koji and Nakamura, Yasunobu},
  year = {2015},
  month = jul,
  journal = {Science},
  volume = {349},
  number = {6246},
  pages = {405--408},
  doi = {10.1126/science.aaa3693}
}

@article{wan2024quantumenhanced,
  title = {Quantum-enhanced metrology in cavity magnonics},
  author = {Wan, Qing-Kun and Shi, Hai-Long and Guan, Xi-Wen},
  year = {2024},
  month = jan,
  journal = {Phys. Rev. B},
  volume = {109},
  number = {4},
  pages = {L041301},
  doi = {10.1103/PhysRevB.109.L041301}
}

@article{wang2016magnon,
  title = {Magnon Kerr effect in a strongly coupled cavity-magnon system},
  author = {Wang, Yi-Pu and Zhang, Guo-Qiang and Zhang, Dengke and Luo, Xiao-Qing and Xiong, Wei and Wang, Shuai-Peng and Li, Tie-Fu and Hu, C.-M. and You, J. Q.},
  year = {2016},
  month = dec,
  journal = {Phys. Rev. B},
  volume = {94},
  number = {22},
  pages = {224410},
  doi = {10.1103/PhysRevB.94.224410}
}

@article{luo2021nonlocal,
  title={Nonlocal magnon entanglement generation in coupled hybrid cavity systems},
  author={Luo, Da-Wei and Qian, Xiao-Feng and Yu, Ting},
  journal={Optics Letters},
  volume={46},
  number={5},
  pages={1073--1076},
  year={2021},
  publisher={Optical Society of America}
}

@article{wolski2020dissipationbased,
  title = {Dissipation-Based Quantum Sensing of Magnons with a Superconducting Qubit},
  author = {Wolski, S. P. and {Lachance-Quirion}, D. and Tabuchi, Y. and Kono, S. and Noguchi, A. and Usami, K. and Nakamura, Y.},
  year = {2020},
  month = sep,
  journal = {Phys. Rev. Lett.},
  volume = {125},
  number = {11},
  pages = {117701},
  doi = {10.1103/PhysRevLett.125.117701}
}

@article{zarerameshti2022cavity,
  title = {Cavity magnonics},
  author = {Zare Rameshti, Babak and Viola Kusminskiy, Silvia and Haigh, James A. and Usami, Koji and {Lachance-Quirion}, Dany and Nakamura, Yasunobu and Hu, Can-Ming and Tang, Hong X. and Bauer, Gerrit E. W. and Blanter, Yaroslav M.},
  year = {2022},
  month = sep,
  journal = {Phys. Rep.},
  volume = {979},
  pages = {1--61},
  doi = {10.1016/j.physrep.2022.06.001}
}

@article{zhang2014strongly,
  title = {Strongly Coupled Magnons and Cavity Microwave Photons},
  author = {Zhang, Xufeng and Zou, Chang-Ling and Jiang, Liang and Tang, Hong X.},
  year = {2014},
  month = oct,
  journal = {Phys. Rev. Lett.},
  volume = {113},
  number = {15},
  pages = {156401},
  doi = {10.1103/PhysRevLett.113.156401}
}

@article{zhang2016cavity,
  title = {Cavity magnomechanics},
  author = {Zhang, Xufeng and Zou, Chang-Ling and Jiang, Liang and Tang, Hong X.},
  year = {2016},
  month = mar,
  journal = {Sci. Adv.},
  volume = {2},
  number = {3},
  pages = {e1501286},
  doi = {10.1126/sciadv.1501286}
}

@article{zhang2016optomagnonic,
  title = {Optomagnonic Whispering Gallery Microresonators},
  author = {Zhang, Xufeng and Zhu, Na and Zou, Chang-Ling and Tang, Hong X.},
  year = {2016},
  month = sep,
  journal = {Phys. Rev. Lett.},
  volume = {117},
  number = {12},
  pages = {123605},
  doi = {10.1103/PhysRevLett.117.123605}
}

@article{zhang2019theory,
  title = {Theory of the magnon Kerr effect in cavity magnonics},
  author = {Zhang, GuoQiang and Wang, YiPu and You, JianQiang},
  year = {2019},
  month = aug,
  journal = {Sci. China-Phys. Mech. Astron.},
  volume = {62},
  number = {8},
  pages = {987511},
  doi = {10.1007/s11433-018-9344-8}
}

@article{zhang2025gain,
  title = {Gain--loss coupled systems},
  author = {Zhang, Chunlei and Kim, Mun and Zhang, Yi-Hui and Wang, Yi-Pu and Trivedi, Deepanshu and Krasnok, Alex and Wang, Jianbo and Isleifson, Dustin and Roshko, Roy and Hu, Can-Ming},
  year = {2025},
  month = mar,
  journal = {APL Quantum},
  volume = {2},
  number = {1},
  pages = {011501},
  doi = {10.1063/5.0250178}
}

@article{zhao2021driven,
  title = {Driven dissipative quantum dynamics in a cavity magnon-polariton system},
  author = {Zhao, Guogan and Wang, Yong and Qian, X.-F.},
  year = {2021},
  month = oct,
  journal = {Phys. Rev. B},
  volume = {104},
  number = {13},
  pages = {134423},
  doi = {10.1103/PhysRevB.104.134423}
}

@article{zhao2025theory,
  title = {Theory of the magnon Purcell effect in a cavity magnonic system},
  author = {Zhao, Guogan and Wang, Yong and Qian, Xiao-Feng},
  year = {2025},
  month = jun,
  journal = {Phys. Rev. B},
  volume = {111},
  number = {21},
  pages = {214428},
  doi = {10.1103/9jw6-w9lw}
}

@article{zheng2023tutorial,
  title = {Tutorial: Nonlinear magnonics},
  author = {Zheng, Shasha and Wang, Zhenyu and Wang, Yipu and Sun, Fengxiao and He, Qiongyi and Yan, Peng and Yuan, H. Y.},
  year = {2023},
  month = oct,
  journal = {J. Appl. Phys.},
  volume = {134},
  number = {15},
  pages = {151101},
  doi = {10.1063/5.0152543}
}

\end{document}